\documentclass[aps,prb,twocolumn,superscriptaddress,showpacs]{revtex4-2}
\usepackage{graphicx}
\usepackage{amsmath}
\usepackage{amssymb}
\usepackage{hyperref}
\usepackage{dcolumn}
\usepackage{bm}
\usepackage[utf8x]{inputenc}
\usepackage{xcolor}
\usepackage{textcomp}
\begin{document}
\title{\texorpdfstring{{Coupled Magnetic and Lattice Excitations in Multiferroic Ba$_{3}$HoRu$_{2}$O$_{9}$: Insights from Neutron Scattering, Raman Spectroscopy, and Machine-Learned Phonon Calculations}}{{Coupled Magnetic and Lattice Excitations in MultiferroicBa$_{3}$HoRu$_{2}$O$_{9}$: Insights from Neutron Scattering, Raman Spectroscopy, and Machine-Learned Phonon Calculations}}}

\author{E. Kushwaha}
\email{ektak22bs@rgipt.ac.in}
\affiliation{Rajiv Gandhi Institute of Petroleum Technology, Jais, Amethi, 229304, India}
\author{S. Ghosh}
\affiliation{Rajiv Gandhi Institute of Petroleum Technology, Jais, Amethi, 229304, India}
\author{G. Roy}
\affiliation{Rajiv Gandhi Institute of Petroleum Technology, Jais, Amethi, 229304, India}
\author{M. Kumar}
\affiliation{Rajiv Gandhi Institute of Petroleum Technology, Jais, Amethi, 229304, India}
\author{M. D.Le}
\affiliation{ISIS Neutron and Muon Source, STFC, Rutherford Appleton Laboratory, Chilton, Oxon OX11 0QX, United Kingdom}
\author{J. Shannigrahi}
\affiliation{School of Physical Sciences, Indian Institute of Technology Goa, Farmagudi, Goa 403401, India}
\author{S.D. Kaushik}
\affiliation{UGC - DAE Consortium for Scientific Research Mumbai Centre, Bhabha Atomic Research Centre, 246 C Common Facility Building, Mumbai-400085, India}
\author{D. T. Adroja}
\affiliation{ISIS Neutron and Muon Source, STFC, Rutherford Appleton Laboratory, Chilton, Oxon OX11 0QX, United Kingdom}
\affiliation{Highly Correlated Matter Research Group, Physics Department, University of Johannesburg, Auckland Park 2006, South Africa}
\author{T. Basu}
\email{tathamay.basu@rgipt.ac.in}
\affiliation{Rajiv Gandhi Institute of Petroleum Technology, Jais, Amethi, 229304, India}

{\begin{abstract}
Inelastic neutron scattering (INS) spectra of correlated $4d$-$4f$ multiferroic oxides are difficult to interpret because magnetic, crystal-field, and lattice excitations frequently overlap in energy. The hexagonal 6H perovskite Ba$_3$HoRu$_2$O$_9$ provides an ideal platform to investigate this interplay owing to the coexistence of Ru$_2$O$_9$ molecular units and localized Ho$^{3+}$ moments. To identify the contributions from these different excitations, we combine INS with linear spin-wave calculations, crystal-field analysis, Raman spectroscopy, and machine-learned force field (MLFF) phonon calculations. A magnetic excitation below 6.2~meV, observed only in the low-temperature magnetic phase, is reproduced by linear spin-wave theory with intradimer Ru-Ru and Ru-Ho exchange of comparable size and a Ho single-ion anisotropy, and is therefore consistent with a collective spin-wave excitation of the coupled Ru-Ho magnetic network. At higher energies, broad excitations centered near 20, 39, 70, and 90~meV are observed that persist far above the magnetic ordering temperatures. The magnetic part of the 39~meV excitation, isolated by subtracting the Bose-corrected high-temperature data, follows the Ho$^{3+}$ magnetic form factor. Crystal-field calculations based on the Stevens formalism place a group of strong Ho$^{3+}$ transitions in this energy range, while Raman spectroscopy and MLFF phonon calculations identify optical phonons with comparable energies. Together, these complementary results show that the broad INS feature near 39~meV arises from overlapping Ho$^{3+}$ crystal-field excitations and lattice vibrations. These findings establish a microscopic framework for understanding the interplay between spin, crystal-field, and lattice degrees of freedom in this multiferroic $4d$-$4f$ compound.
\end{abstract}}

\maketitle

\section{Introduction}

Correlated transition-metal oxides exhibit a wide variety of emergent phenomena arising from the interplay among charge, spin, orbital, and lattice degrees of freedom. In these materials, competing interactions between electron correlations, spin-orbit coupling (SOC), crystal-electric field (CEF), and complex exchange interactions often stabilize unconventional magnetic ground states~\cite{witczak2014correlated}, including frustrated magnetism~\cite{ishiguro2013dynamical}, molecular magnetism~\cite{streltsov2016covalent,ziat2017frustrated}, orbital-selective Mott states~\cite{chen2020realization}, multiferroicity~\cite{kimura2003magnetic,hur2004electric,chapon2004structural}, and exotic spin textures. Understanding the elementary excitations associated with these competing interactions is essential for establishing microscopic models of correlated electron systems and for developing multifunctional materials with potential applications in spintronics and magnetoelectric devices.

Among correlated oxides, ruthenium-based compounds occupy a unique position because the spatially extended $4d$ orbitals give rise to an intricate balance between electronic itinerancy, electron correlation and SOC~\cite{cava2004schizophrenic,witczak2014correlated}. Compared with their $3d$ counterparts, Ru-based oxides possess stronger metal-ligand hybridization and relatively strong SOC, leading to a rich variety of magnetic and electronic ground states that are highly sensitive to crystal structure and exchange geometry~\cite{cava2004schizophrenic,streltsov2016covalent}. Consequently, ruthenates provide an excellent platform for investigating the evolution from localized to itinerant magnetism as well as the emergence of unconventional magnetic excitations.

The situation becomes considerably more complex when magnetic rare-earth ($R$) ions are incorporated into the lattice. The localized $4f$ electrons introduce CEF excitations that coexist and often overlap with magnetic excitations of the transition-metal sublattice and optical phonons, complicating the interpretation of inelastic neutron spectra~\cite{ziat2017frustrated,adroja2020muon,roy2026quasiparticle}. Previous studies on pyrochlore and double-perovskite ruthenates have demonstrated that strong Ru-$R$ interactions can substantially modify the magnetic ground state and generate unusual spin dynamics. For example, in Ho$_2$Ru$_2$O$_7$, the Ru-induced internal field strongly modifies the Ho$^{3+}$ crystal-field states via the Ru-Ho exchange, leading to an unconventional partially ordered Ho state with residual entropy~\cite{gardner2005spin,museur2026ferromagnetic}. Ba$_2$DyRuO$_6$ exhibits a collinear, Ising-like antiferromagnetic ground state in which the Ru$^{5+}$ and Dy$^{3+}$ moments order simultaneously below $T_N = 47$~K, driven by strong $4d$-$4f$ exchange interactions~\cite{roy2026quasiparticle}.
\begin{figure}[ht!]
\centering
\includegraphics[width= \columnwidth,angle=0]{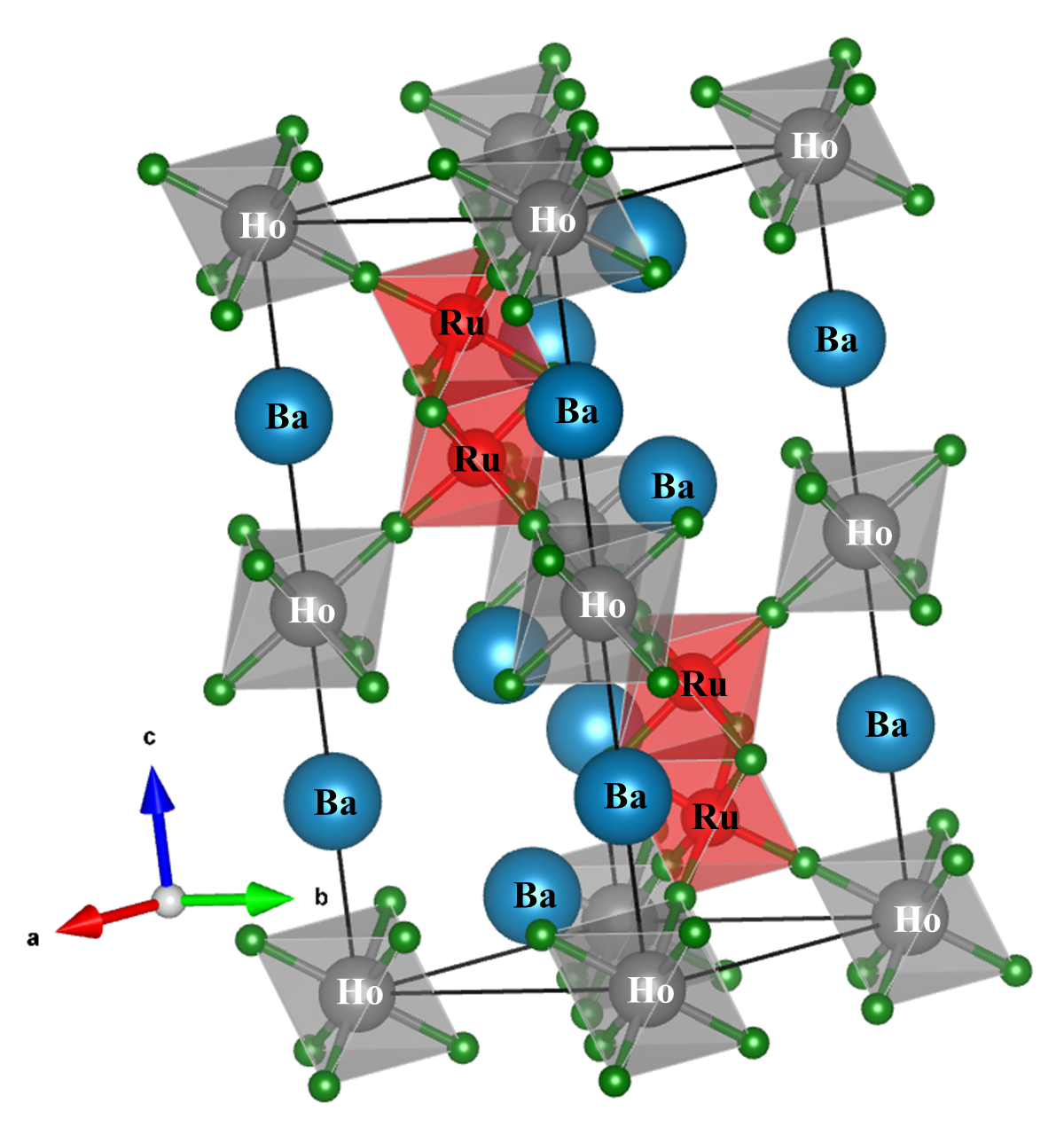}
\caption{\label{Fig-1} Crystal structure of the hexagonal 6H-perovskite Ba$_{3}$HoRu$_{2}$O$_{9}$ , showing face-sharing Ru$_{2}$O$_{9}$ dimers connected through corner-sharing HoO$_{6}$ octahedra. The geometry supports competing Ru-Ru and Ru-Ho magnetic interactions.}
\end{figure}

The hexagonal 6H-perovskite family Ba$_3R$Ru$_2$O$_9$ ($R$ = rare earth or nonmagnetic ion) is an especially attractive class of materials for exploring these competing interactions. This series shows magnetodielectric behaviour, revealing strong spin-dipole coupling~\cite{basu2018magnetodielectric,basu2019enhancement}, and spin-driven ferroelectricity has been demonstrated in Ba$_3$HoRu$_2$O$_9$, for the first time in a $4d$-$4f$ correlated system, indicating strong spin-lattice coupling~\cite{basu2020cooperative,kushwaha2024origin}. The crystal structure consists of face-sharing Ru$_2$O$_9$ dimers connected through corner-sharing $R$O$_6$ octahedra (Fig. 1), producing a lattice in which direct Ru-Ru interactions coexist with long-range Ru-O-$R$ superexchange pathways~\cite{doi2002magnetic,doi2001crystal}. Each Ru ion in these dimers has an average valence of approximately +4.5, yielding seven $4d$ electrons per dimer~\cite{ziat2017frustrated,senn2013spin}. The competition between intradimer electron hopping, intra- and interdimer exchange interactions, and coupling to the rare-earth moments gives rise to a rich variety of electronic and magnetic ground states~\cite{ziat2017frustrated,chen2020realization,chen2019large,hayashida2025magnetic,basu2020cooperative}. The Ba$_3R$Ru$_2$O$_9$ family also exhibits an intriguing magnetocaloric effect (MCE) with magnetocaloric switching for the $R$ = Ho, Gd, and Tb members, highlighting the strong coupling between their magnetic and electronic degrees of freedom~\cite{kumar2026intriguing}.

Previous neutron scattering studies on compounds containing nonmagnetic ions ($R$ = In, Y, Lu) have revealed localized molecular excitations associated with the Ru$_2$O$_9$ dimers, demonstrating that these excitations are an intrinsic feature of this structural family~\cite{ziat2017frustrated}. Replacing the nonmagnetic ion with a magnetic rare-earth ion introduces an additional magnetic degree of freedom that can interact with the Ru$_2$O$_9$ dimers, leading to diverse magnetic ground states~\cite{basu2019enhancement,basu2020cooperative,kushwaha2025unconventional}. Long-range antiferromagnetic order has been reported in several members of this family, including those with nonmagnetic La and Y~\cite{senn2013spin}, whereas Ba$_3$NdRu$_2$O$_9$ exhibits ferromagnetic order linked to the more itinerant nature of the Nd $4f$ states~\cite{senn2013spin,basu2018magnetodielectric}. Interestingly, cooperative magnetic ordering of both Ru and the rare earth has been reported for the heavy rare-earth members Ho and Tb, owing to strong $4d$-$4f$ coupling~\cite{basu2020cooperative,kushwaha2025unconventional}.

Ba$_3$HoRu$_2$O$_9$ is a member of this family. Neutron diffraction measurements have shown that this compound undergoes two successive magnetic transitions: an antiferromagnetic transition at $T_{N1} \approx 50$~K with propagation vector $\mathbf{k}_1 = (1/2, 0, 0)$, followed by a second magnetic transition below $T_{N2} \approx 10$~K characterized by $\mathbf{k}_2 = (1/4, 1/4, 0)$~\cite{basu2020cooperative}. The low-temperature phase is accompanied by spin-driven ferroelectricity through the inverse Dzyaloshinskii-Moriya (DM) interaction arising from the noncollinear magnetic structure, making Ba$_3$HoRu$_2$O$_9$ an attractive model system for studying the coupling between magnetic and lattice degrees of freedom~\cite{kushwaha2024origin}. Despite these intriguing magnetic and multiferroic properties, inelastic neutron scattering investigations of the magnetic members of this lanthanide series remain largely unexplored. Such studies are therefore needed to understand the microscopic origin of the low- and high-energy excitations and the correlations between spin, phonon and CEF in this spin-driven ferroelectric system.

In the present work, we investigate the magnetic and lattice excitations of Ba$_3$HoRu$_2$O$_9$ using inelastic neutron scattering (INS) over a wide range of energy and temperature. To unravel the microscopic origin of these excitations, the neutron scattering measurements are complemented by linear spin-wave calculations, CEF calculations within the Stevens formalism, Raman spectroscopy, and machine-learned force field (MLFF) phonon calculations. This combined experimental and theoretical approach provides a comprehensive framework for understanding the elementary magnetic and lattice excitations in this complex $4d$-$4f$ oxide.

\section{Experimental details}

High-quality polycrystalline Ba$_{3}$HoRu$_{2}$O$_{9}$ was synthesized by a conventional solid-state reaction using stoichiometric amounts of BaCO$_{3}$ (99.99\%), Ho$_{2}$O$_{3}$ (99.99\%), and RuO$_{2}$ (99.99\%), as described in Refs.~\cite{basu2019enhancement,doi2002magnetic}. INS measurements were performed on the MARI time-of-flight spectrometer at the ISIS Neutron and Muon Source (UK)~\cite{le2023upgrade}. Approximately 3.8~g of powder was mounted in an annular geometry in a thin Al-foil envelope inside a thin-walled aluminium can. The sample was cooled in He exchange gas in a $^{4}$He cryostat down to 1.8~K. The measurements were made in repetition-rate multiplication (RRM) mode with a primary incident energy $E_i = 160$~meV and the Gd Fermi chopper at 400~Hz, which also gave data for $E_i = 11.3$ and 28.3~meV in the same measurement. This covers the low-energy magnetic excitations as well as the high-energy crystal-field and lattice excitations. Data were collected at 1.8, 30, 100, and 300~K, spanning both magnetically ordered phases and the paramagnetic regime. Corrections for detector efficiency and background were applied using the Mantid and DAVE software packages~\cite{arnold2014mantid,azuah2009dave}, and the Bose population factor was applied where stated.

Linear spin-wave calculations were performed with the SpinW package~\cite{toth2015linear}. The experimentally determined magnetic structure and crystallographic parameters~\cite{basu2020cooperative} were used as input, and symmetry-allowed exchange pathways were explored to reproduce the observed INS spectra. Crystal-electric-field calculations for Ho$^{3+}$ were carried out within the Stevens operator formalism~\cite{stevens1952matrix,hutchings1964point}. The initial crystal-field parameters were obtained from the point-charge model of the Mantid CrystalField package~\cite{arnold2014mantid}, based on the experimentally refined crystal structure. The crystal-field Hamiltonian was diagonalized to obtain the energy levels, wave functions, and neutron transition intensities, and its overall energy scale was adjusted by a uniform scaling of the Stevens parameters for comparison with the INS spectrum. Lattice dynamics were modelled with the MatterSim machine-learned force field (MLFF)~\cite{yang2024mattersim} within the INSPIRED framework~\cite{cheng2023direct}, which gives the phonon dispersion and the powder-averaged neutron scattering intensity for direct comparison with experiment. Raman spectra were measured at room temperature with a Princeton Instruments Acton SpectraPro-2500i spectrometer, using a 532~nm diode-pumped solid-state laser (Laser Quantum GEM) operated at 50~mW. The spectral resolution was approximately 1.0~cm$^{-1}$, with a wavelength calibration stability better than 0.5~cm$^{-1}$. The $\Gamma$-point phonon modes and eigenvectors used for the Raman assignment were calculated with phonopy and the MACE-MP-0 force field~\cite{batatia2025foundation,togo2015first}.

\section{Results and discussion}
\subsection{Low-energy magnetic excitations}

To investigate the collective magnetic excitations in Ba$_3$HoRu$_2$O$_9$, INS measurements were performed below and above the magnetic ordering temperatures with an incident neutron energy $E_i = 28.3$~meV. Fig. 2(a) shows the INS intensity as a function of energy transfer and momentum transfer $Q$ at 1.8~K, which reveals a broad excitation below 6.2~meV. The excitation has its maximum intensity at low momentum transfer and gradually loses spectral weight with increasing ($Q$) (Supplemental Information~\cite{SI}, Fig. S1), consistent with its magnetic origin. To quantify the excitation spectrum, we extracted the momentum-integrated INS intensity ($0.5 \le |Q| \le 3.5$~\AA$^{-1}$) as a function of energy transfer for temperatures between 1.8 and 300~K [Fig. 2(b)]. s temperature increases above $T_{N2}$ this excitation vanishes, and no additional low-energy magnetic excitations are observed within the experimental resolution. However, a weak feature near 12 meV becomes visible only at 300 K, suggesting a phonon or thermally activated contribution. The excitation below 6.2~meV is therefore associated with the low-temperature magnetic phase below $T_{N2}$, in which the Ho moment reaches its full value~\cite{basu2020cooperative}.Its finite bandwidth, its $Q$ dependence and its disappearance above $T_{N2}$ indicate that it is a spin-wave excitation rather than an isolated intradimer molecular excitation.
\begin{figure}[!ht]
\centering
\includegraphics[width=3.5in,angle=0]{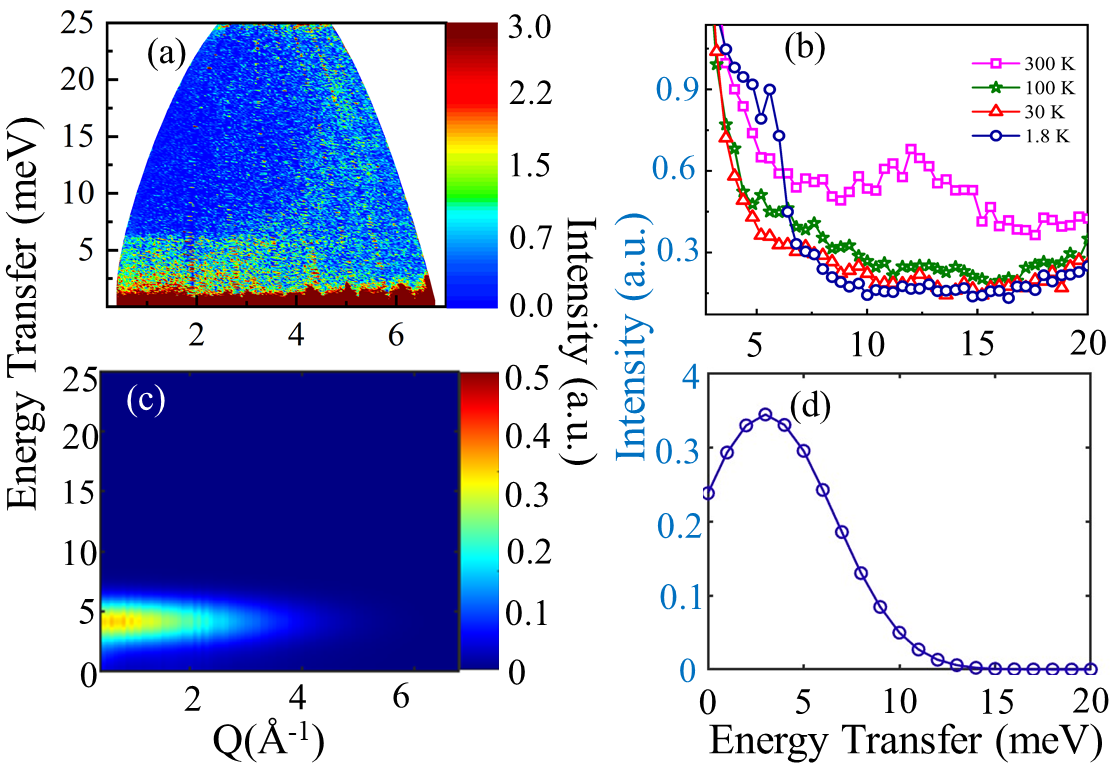}
\caption{\label{Fig-2} (a) Color-coded map of the inelastic neutron scattering intensity of Ba$_3$HoRu$_2$O$_9$ as a function of momentum transfer ($|Q|$) and energy transfer, measured at 1.8 K with an incident neutron energy of $E_i = 28.3$ meV. (b) Temperature dependence of the $|Q|$-integrated inelastic neutron scattering intensity as a function of energy transfer for temperatures between 1.8 and 300 K, integrated over the momentum-transfer range $0.5 \leq |Q| \leq 3.5$~\AA$^{-1}$. (c) Simulated color-coded contour map of the calculated spin-wave excitation spectrum as described in text as a function of momentum transfer $|Q|$ and energy transfer and (d) Intensity vs Energy transfer for the simulated 1D cut obtained from SpinW for $|Q| = 0.5$ to $3.5$~\AA$^{-1}$.}
\end{figure}
To interpret the microscopic origin of this excitation, we modelled the spin dynamics using linear spin-wave theory within the SpinW package~\cite{toth2015linear}.  Neutron diffraction shows that both the Ru and Ho sublattices carry ordered magnetic moments down to the lowest measured temperature (1.5~K)~\cite{basu2020cooperative}. Both ions were therefore included as magnetic sites in the model, using the experimentally determined lattice constants and the Wyckoff positions of Ru and Ho. The spin Hamiltonian is
\begin{equation}
H = \sum_{ij} J_{ij}\, \mathbf{S}_i \cdot \mathbf{S}_j + \sum_i D_i \left(S_i^z\right)^2 ,
\end{equation}
where $J_{ij}$ is the exchange interaction between the $i$th and $j$th spins $\mathbf{S}_i$ and $\mathbf{S}_j$, and $D_i$ is the single-ion anisotropy.The anisotropy term is applied only to the Ho atoms and is used as an effective parameter of the spin model.

\begin{table}[t]
\caption{\label{tab:exchange}Exchange interactions obtained from SpinW (negative values for FM and positive values for AFM coupling).}
\begin{ruledtabular}
\begin{tabular}{lcc}
Exchange interaction & Bond length (\AA) & Value (meV) \\
\hline
$J_1$ (Ru-Ru, intradimer) & 2.55 & 1.57 \\
$J_2$ (Ru-Ho) & 4.11 & 1.45 \\
$J_3$ (Ru-Ru, interdimer) & 5.76 & $-0.05$ \\
$J_4$ (Ho-Ho) & 5.85 & $-0.05$ \\
\end{tabular}
\end{ruledtabular}
\end{table}

We optimized the exchange parameters in two steps. First, we varied them until the calculated ground state reproduced the magnetic structure determined by neutron diffraction~\cite{basu2020cooperative}. Then, the excitation calculated with these parameters was compared with the INS data. The optimized exchange constants are listed inTable I. $J_1$ (intradimer Ru-Ru) and $J_2$ (Ru-Ho) are much larger than $J_3$ (interdimer Ru-Ru) and $J_4$ (Ho-Ho), although all four exchange interactions are required to reproduce the experimentally determined magnetic structure. Removing either $J_3$ or $J_4$ changes the calculated spin arrangement, indicating that these weaker interactions also contribute to selecting the ground state. The magnetic structures obtained with all four interactions and with individual exchange terms removed are presented in Fig. S2. The excitation calculated with these exchange interactions reproduces the overall shape of the observed spectrum, but the calculated energies are too low (Fig. S3(a)). Introducing a Ho-site anisotropy $D_{zz} = -0.25$~meV brings the energy into agreement with experiment without changing the magnetic structure (Fig. S3(b)), and gives the spectrum as shown in Fig. 2(c,d). The exchange interactions therefore determine the magnetic structure, while $D_{zz}$ sets the energy scale of the excitation. Since both $J$ and $D_{zz}$ change the energy, their individual values are not unique; the result that is fixed by the data is the hierarchy $J_1, J_2 \gg |J_3|, |J_4|$.
To see the roles of $J_1$ and $J_2$, we calculated the excitation with only one of them (Fig. S4). With $J_1$ only, which describes isolated Ru$_2$O$_9$ dimers, the intensity is weak and restricted to energies below about 3~meV. With $J_2$ only, a dispersive excitation reaching about 5~meV appears, on the energy scale of the observed one. The observed spectrum (Fig. 2(c,d)) is reproduced only when both interactions are included. Although $J_2$ is large, no dispersion is resolved in the measured band because of the powder averaging, which averages the dispersion over all directions of $Q$. In the calculation the mode disperses only over a narrow energy range: its powder-averaged intensity is centred at 3.9~meV with an rms spread of 0.9~meV (FWHM $\approx$ 2~meV). Together with the instrumental resolution (FWHM $\approx$ 0.9~meV at the elastic line), this gives the broad band between about 2 and 6~meV in Fig. 2. The excitation below 6.2~meV thus arises from collective spin waves of the exchange-coupled Ru-Ho network and not from isolated Ru$_2$O$_9$ dimers. The optimized exchange parameters reveal a pronounced hierarchy of magnetic interactions, with the largest values for the intradimer Ru-Ru exchange ($J_1$) and the Ru-Ho exchange ($J_2$), whereas the interdimer Ru-Ru ($J_3$) and Ho-Ho ($J_4$) interactions are comparatively weak. This hierarchy indicates that the magnetic properties of Ba$_3$HoRu$_2$O$_9$ are governed mainly by Ru$_2$O$_9$ dimers coupled through the Ho sublattice, while the interdimer Ru-Ru and Ho-Ho exchange interactions play a secondary role. Neutron diffraction shows that the magnetic ground state cannot be understood in terms of isolated Ru$_2$O$_9$ dimers, both Ru and Ho moments develop long-range order simultaneously below $T_{N1} \approx 50$~K, which indicates a strong Ru($4d$)-Ho($4f$) coupling~\cite{basu2020cooperative}. In our model the Ru-Ho exchange $J_2$ is comparable to the intradimer exchange $J_1$ and is therefore not a small perturbation. It is mediated by Ru-O-Ho superexchange pathways with a Ho-O-Ru angle of 178.6$^\circ$, and such a nearly linear path is expected to favour antiferromagnetic coupling according to the Goodenough-Kanamori-Anderson rules~\cite{Goodenough1955,Kanamori1959,Anderson1959}. A Ru-Ho exchange comparable to $J_1$ is therefore consistent with the simultaneous ordering of the Ru and Ho sublattices observed by neutron diffraction~\cite{basu2020cooperative}.

The much weaker interdimer Ru-Ru ($J_3$) and Ho-Ho ($J_4$) interactions do not set the energy of the excitation, but they select the observed spin arrangement. In particular, the Ho-Ho exchange ($J_4 = -0.05$~meV, about 0.6~K) is far too small to order the Ho moments at 50~K, which shows that the Ho moments are polarized by the Ru sublattice through the Ru-Ho exchange $J_2$. These weaker interactions are also likely to influence the detailed magnetic configuration and the low-temperature phase competition below $T_{N2}$, where neutron diffraction reveals the coexistence of two magnetic propagation vectors~\cite{basu2020cooperative}. The emergence of this second magnetic phase is therefore likely driven by the competition between the stronger Ru-Ru and Ru-Ho exchange interactions together with the magnetic anisotropy, rather than by the weak interdimer interactions alone.

\subsection{High-energy excitations}

To further investigate the high-energy excitation spectrum, INS measurements were performed with an incident neutron energy of 160~meV. Fig. 3(a-d) shows the INS spectra collected at 1.8, 30, 100 and 300~K, respectively. Since magnetic and phonon scattering have distinct momentum-transfer ($Q$) dependences, the spectra were analysed separately over two $Q$ ranges, $1 < |Q| < 8$~\AA$^{-1}$ and $8 < |Q| < 16$~\AA$^{-1}$, and the corresponding momentum-integrated spectra are presented in Fig. 3(e,f). Broad excitations are observed at approximately 20, 39, 70, and 90~meV in both $Q$ ranges. Unlike the low-energy collective mode, these excitations remain clearly visible over the entire temperature range from 1.8 K to 300~K, indicating that they are not directly associated with long-range magnetic order.
\begin{figure}[t]
\centering
\includegraphics[width=\columnwidth]{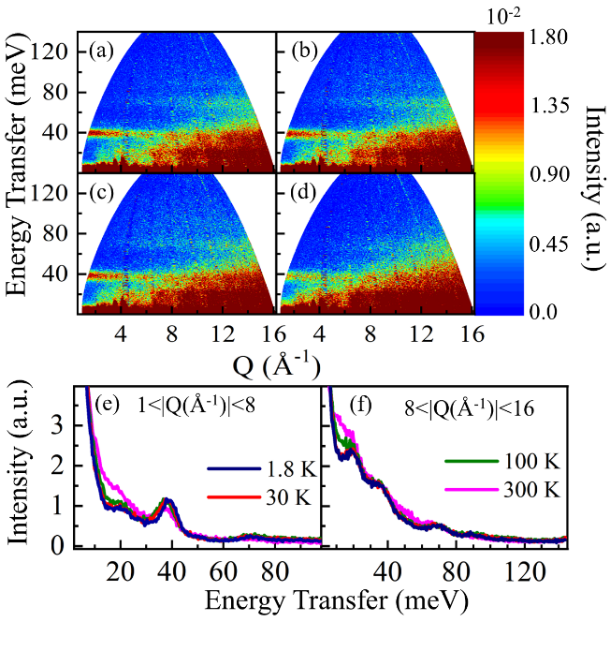}
\caption{\label{fig:highE}INS intensity of Ba$_3$HoRu$_2$O$_9$ as a function of energy and momentum transfer, measured with $E_i = 160$~meV at (a) 1.8~K, (b) 30~K, (c) 100~K, and (d) 300~K. Momentum-integrated intensity as a function of energy transfer between 1.8 and 300~K for (e) $1 < |Q| < 8$~\AA$^{-1}$ and (f) $8 < |Q| < 16$~\AA$^{-1}$.}
\end{figure}

Among these, the excitation near 39~meV carries the largest spectral weight at all temperatures. We therefore analysed its thermal evolution by fitting the momentum-integrated spectra ($1 \le |Q| \le 8$~\AA$^{-1}$) with a Gaussian function to extract the peak position, linewidth, and integrated intensity. Representative fits for 1.8 and 300~K are shown in Fig. S5~\cite{SI}. The peak position is $39.09 \pm 0.06$~meV at 1.8~K and $37.25 \pm 0.23$~meV at 300~K. At the same time, the full width at half maximum (FWHM) increases from $6.58 \pm 0.13$~meV to $8.19 \pm 0.55$~meV, and the integrated intensity decreases by about 28\%. The excitation nevertheless remains clearly observable at 300~K, well above both magnetic ordering temperatures. This weak temperature dependence and the persistence of the excitation far above the ordering temperatures are inconsistent with a conventional spin-wave excitation associated with long-range magnetic order. The temperature dependence alone, however, cannot identify its microscopic origin, because localized crystal-electric-field excitations, molecular magnetic excitations of the Ru$_2$O$_9$ dimers, and optical phonons may all survive well into the paramagnetic regime.

To further characterize these excitations, we examined the momentum dependence of their intensity (Fig. S6). For the peaks between 16-22, 60-74, and 85-95~meV (Fig. S6(a,c,d)), the intensity increases gradually with increasing $|Q|$. This behaviour is characteristic of phonon scattering and shows that these excitations are mainly lattice vibrations. The excitation near 39~meV (Fig. S6(b)) behaves differently, its intensity is strongest at small $|Q|$, which points to a magnetic contribution, but it rises again above about 5~\AA$^{-1}$, showing that phonons also contribute. To separate the magnetic and phonon contributions, we performed a Bose-factor correction, in which the 300~K spectrum was divided by $n(\omega)+1$ and subtracted from the 1.8~K spectrum. After this subtraction, only the magnetic part of the 39~meV feature remains [Fig.~S6(e)]. Between 1.4 and 5.5~\AA$^{-1}$, this magnetic intensity follows the Ho$^{3+}$ form factor~\cite{Brown2004}, compared with its value at 1.7~\AA$^{-1}$, about 70\% remains at 3~\AA$^{-1}$ and 50\% at 4~\AA$^{-1}$, close to the 75\% and 55\% expected for Ho$^{3+}$. In contrast, a Ru$_2$O$_9$ dimer excitation, with intensity given by the Ru form factor multiplied by $[1 - \sin(Qd)/(Qd)]$ ($d = 2.55$~\AA)~\cite{FurrerGudel1979,FurrerWaldmann2013}, would drop to 15\% at 3~\AA$^{-1}$ and 3\% at 4~\AA$^{-1}$, which does not match our data. The details of the Bose-factor correction and the Ho$^{3+}$ form-factor calculation are given in the Supplemental Material~\cite{SI}. Above about 5.5~\AA$^{-1}$, the subtraction is less reliable because some phonon intensity remains. The 39~meV excitation is therefore a superposition of phonon and magnetic scattering, and its magnetic part comes mainly from Ho$^{3+}$. The coexistence of magnetic Ho$^{3+}$ ions, Ru$_2$O$_9$ molecular units, and lattice vibrations in Ba$_3$HoRu$_2$O$_9$ makes it necessary to identify these contributions in more detail. To address this, crystal-electric-field calculations, Raman spectroscopy, and machine-learned phonon calculations were performed, as discussed in the following sections.

\subsection{Crystal-electric-field analysis of Ho$^{3+}$ excitations}\label{sec:cef}

The presence of magnetic Ho$^{3+}$ ions in Ba$_3$HoRu$_2$O$_9$ naturally raises the possibility that localized CEF excitations contribute to the high-energy INS spectrum. In particular, the broad excitation centered near 39~meV persists far above the magnetic ordering temperatures, and its magnetic part follows the Ho$^{3+}$ form factor (Fig. S6(e)). To examine Ho$^{3+}$ CEF transitions, crystal-field calculations were performed within the Stevens operator formalism~\cite{stevens1952matrix,hutchings1964point}. The calculations were based on the experimentally refined crystal structure, in which the Ho$^{3+}$ ion occupies the Wyckoff 2a site of the hexagonal $P6_3/mmc$ structure with local $D_{3d}$ ($\bar{3}m$) point symmetry~\cite{kroumova2003bilbao}. Under this symmetry only six independent Stevens parameters are allowed, and the crystal-field Hamiltonian is expressed as
\begin{equation}
H_{\mathrm{CEF}} = B_2^0 O_2^0 + B_4^0 O_4^0 + B_4^3 O_4^3 + B_6^0 O_6^0 + B_6^3 O_6^3 + B_6^6 O_6^6,
\end{equation}
where $B_k^q$ are the Stevens crystal-field parameters and $O_k^q$ are the corresponding Stevens operator equivalents. The initial parameters obtained from the point-charge model of the Mantid CrystalField package~\cite{arnold2014mantid} are listed in S.I. Table S1 ~\cite{SI}. Diagonalization of the CEF Hamiltonian splits the Ho$^{3+}$ ($J = 8$) multiplet into seventeen crystal-field states, which form eleven distinct levels, singlets and doublets. The neutron-scattering intensity of each CEF transition is determined by the square of the magnetic dipole matrix elements between the initial and final crystal-field wavefunctions, weighted by the thermal (Boltzmann) population of the initial state~\cite{dun2021effective}. The calculated transition energies and relative intensities are summarized in Table S2 and Fig. S7~\cite{SI}, which differ by more than two orders of magnitude, even for levels that are relatively close in energy. This behavior can be understood from the local symmetry of the Ho site. The HoO$_6$ octahedron is almost regular. The Ho-O distance is 2.176~\AA{} and the Ho-O bonds make an angle of 54.62$^\circ$ with the $c$ axis, compared with 54.74$^\circ$ for an ideal octahedron. The $D_{3d}$ Hamiltonian is therefore, to within 0.8~\%, a cubic crystal field with a threefold axis along $c$ plus a small trigonal $B_2^0$ term (Table S3). The crystal field at the Ho site is therefore close to cubic, and the trigonal term only lifts the degeneracy of the cubic levels by a small amount and makes transitions weakly allowed that are forbidden in the cubic limit. Not all symmetry-allowed CEF excitations are therefore expected to have comparable experimental intensity, and the observation of only the most intense excitation is consistent with the calculated variation in transition probabilities and the finite experimental resolution. The initial point charge model predicts the strongest transitions between 25-30~meV, whereas comparatively weaker transitions occur near 35-42~meV. This is inconsistent with the INS measurements where the dominant excitation is observed near 39~meV, suggesting that the point-charge approximation underestimates the overall crystal-field strength.
\begin{figure}[t]
\centering
\includegraphics[width=0.85\columnwidth]{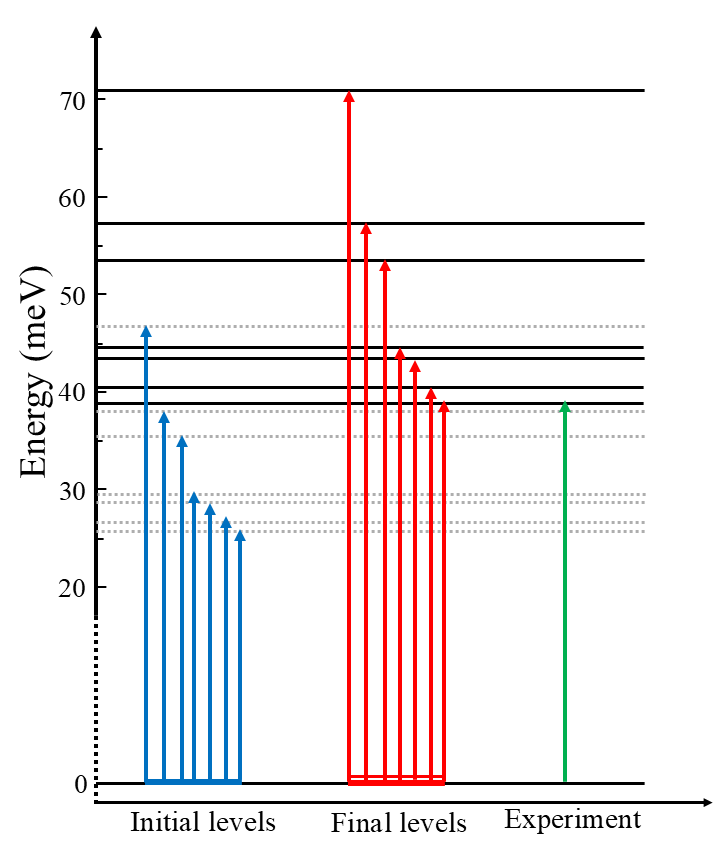}
\caption{\label{fig:levels}Ho$^{3+}$ crystal-field levels from the initial point-charge model (blue) and after uniform scaling of the Stevens parameters by 1.51 (red), compared with the observed excitation (green). Arrows mark transitions from the ground state.}
\end{figure}

\begin{figure}[t]
\centering
\includegraphics[width=0.9\columnwidth]{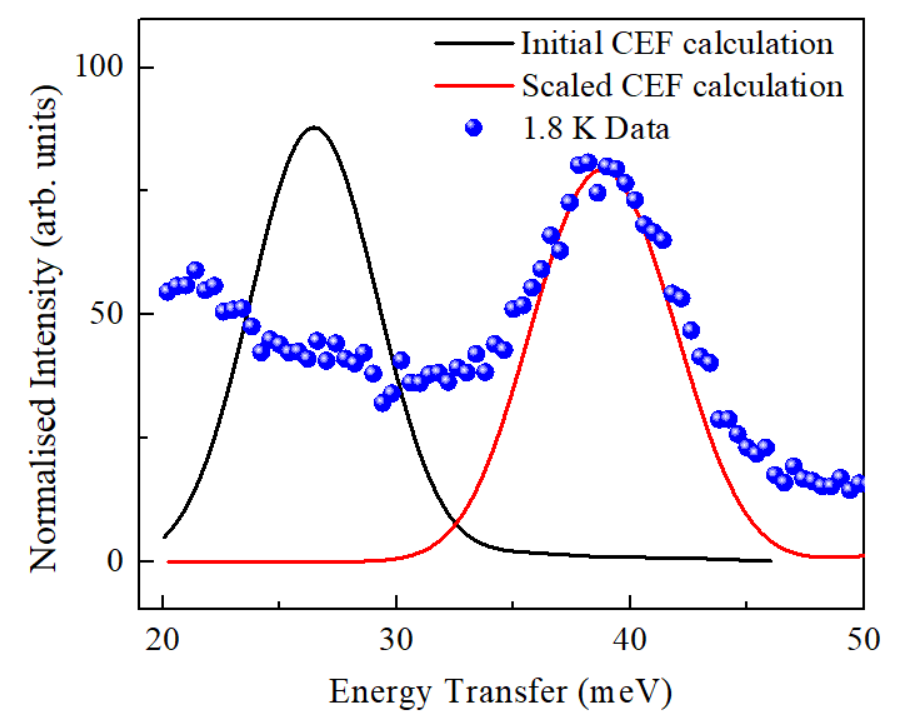}
\caption{\label{fig:cefcomp}INS spectrum at 1.8~K ($1 \le |Q| \le 8$~\AA$^{-1}$) compared with the crystal-field spectra calculated with the initial point-charge parameters (black) and the uniformly scaled Stevens parameters (red). The scaled calculation reproduces the energy of the excitation near 39~meV, whereas the initial point-charge model places the strongest transitions near 26~meV.}
\end{figure}

In systems where several CEF transitions are experimentally resolved, the Stevens parameters are typically refined by simultaneously fitting the transition energies and intensities~\cite{roy2026quasiparticle,hutchings1964point,dun2021effective}. Such a refinement is not feasible for the present system because the Ho$^{3+}$ ion in $D_{3d}$ symmetry is described by six independent Stevens parameters, and even the cubic limit has two ($B_4$ and $B_6$), while the present INS measurements provide only one prominent excitation near 39~meV that can be associated with a Ho$^{3+}$ CEF transition. Moreover, the absolute energy scale of a point-charge calculation is not well defined, because it depends on the radial integrals, their shielding by the outer electrons and the convergence of the lattice sum~\cite{hutchings1964point,dun2021effective,newman2000crystal}. In the Mantid calculation the lattice sum is taken over a finite sphere, for which the second-order term converges poorly. We therefore multiplied all Stevens parameters by a common factor of 1.51 (Table S4), chosen so that the lowest line of the group of strong transitions lies at the observed peak position of 39~meV. The complete scaling analysis is presented in Table S5, while the dependence of this transition energy on the scaling factor is shown in Fig. S8~\cite{SI}. Because the factor is common, the ratios between the parameters, the eigenvectors and the relative transition probabilities are unchanged. This is a phenomenological adjustment of the energy scale to the one observed peak and not an independent refinement~\cite{roy2026quasiparticle,hutchings1964point,dun2021effective}. The resulting scaled energy levels and corresponding transition probabilities are summarized in Table S6 and Fig. S9~\cite{SI}, and the energy-level diagram is shown in Fig. 4. In the scaled model the strongest transitions form a group of four lines at 39.0, 40.6, 43.5 and 44.6~meV, while the next transitions, near 54 and 57~meV, are about ten times weaker. A direct comparison between the measured INS spectrum at 1.8~K and the calculated crystal-field spectra obtained using the initial point-charge model and the uniformly scaled Hamiltonian is presented in Fig. 5.

This analysis indicates that the principal disagreement between the point-charge calculation and the experimental INS spectrum arises from an underestimation of the overall crystal-field energy scale rather than from the symmetry of the crystal-field Hamiltonian.

The calculated CEF transitions also provide a possible explanation for the observed linewidth. The four strongest transitions are distributed over an energy range of approximately 5.6~meV, which is comparable to the experimental FWHM of 6.6~meV. In addition, phonon contributions in the same energy range (Figs. 6 and 8) may further broaden the measured feature.

In the ordered state the exchange field arising from the Ru sublattice may also lift the degeneracy of the Ho$^{3+}$ CEF levels. A quantitative description of this effect would require incorporating the Ru-Ho exchange interaction into the CEF model, which is beyond the present analysis. Localized magnetic excitations associated with Ru$_2$O$_9$ dimers have previously been reported at 31.5-34~meV in the nonmagnetic members Ba$_3M$Ru$_2$O$_9$ ($M$ = In, Y, Lu)~\cite{ziat2017frustrated} and around 35~meV in Ba$_3$LaRu$_2$O$_9$~\cite{chen2020realization}. Since the Ru valence in Ba$_3$HoRu$_2$O$_9$ is the same as in the In-, Y-, and Lu-based compounds, a similar dimer excitation may also contribute to the spectrum. However, its contribution to the feature near 39~meV appears to be limited. In particular, the magnetic intensity follows the Ho$^{3+}$ magnetic form factor rather than the expected momentum dependence of a Ru$_2$O$_9$ dimer excitation (Fig. S6(e)). These results suggest that the magnetic contribution to the 39~meV feature arises predominantly from the Ho$^{3+}$ CEF transitions.

\subsection{Raman spectroscopy}

Room-temperature Raman spectroscopy was performed to probe the zone-centre phonons and to compare their energies with the high-energy excitations observed in the INS measurements. The Raman spectrum shown in Fig. 6 was recorded between 100 and 1200~cm$^{-1}$ and exhibits several well-defined phonon modes at approximately 107, 164, 231, 325, 370, 574, and 774~cm$^{-1}$. Using $E$~(meV)~$= 0.12398 \times \omega$~(cm$^{-1}$), these phonon modes correspond to energies around 13.3, 20.3, 28.6, 40.3, 45.9, 71.2 and 96.0~meV, respectively. Raman spectroscopy probes only the Brillouin-zone-centre ($\Gamma$-point) excitations, and for the $P6_3/mmc$ space group the Raman-active modes are $5A_{1g} + 6E_{1g} + 8E_{2g}$~\cite{kroumova2003bilbao}.
\begin{figure}[t]
\centering
\includegraphics[width=0.8\columnwidth]{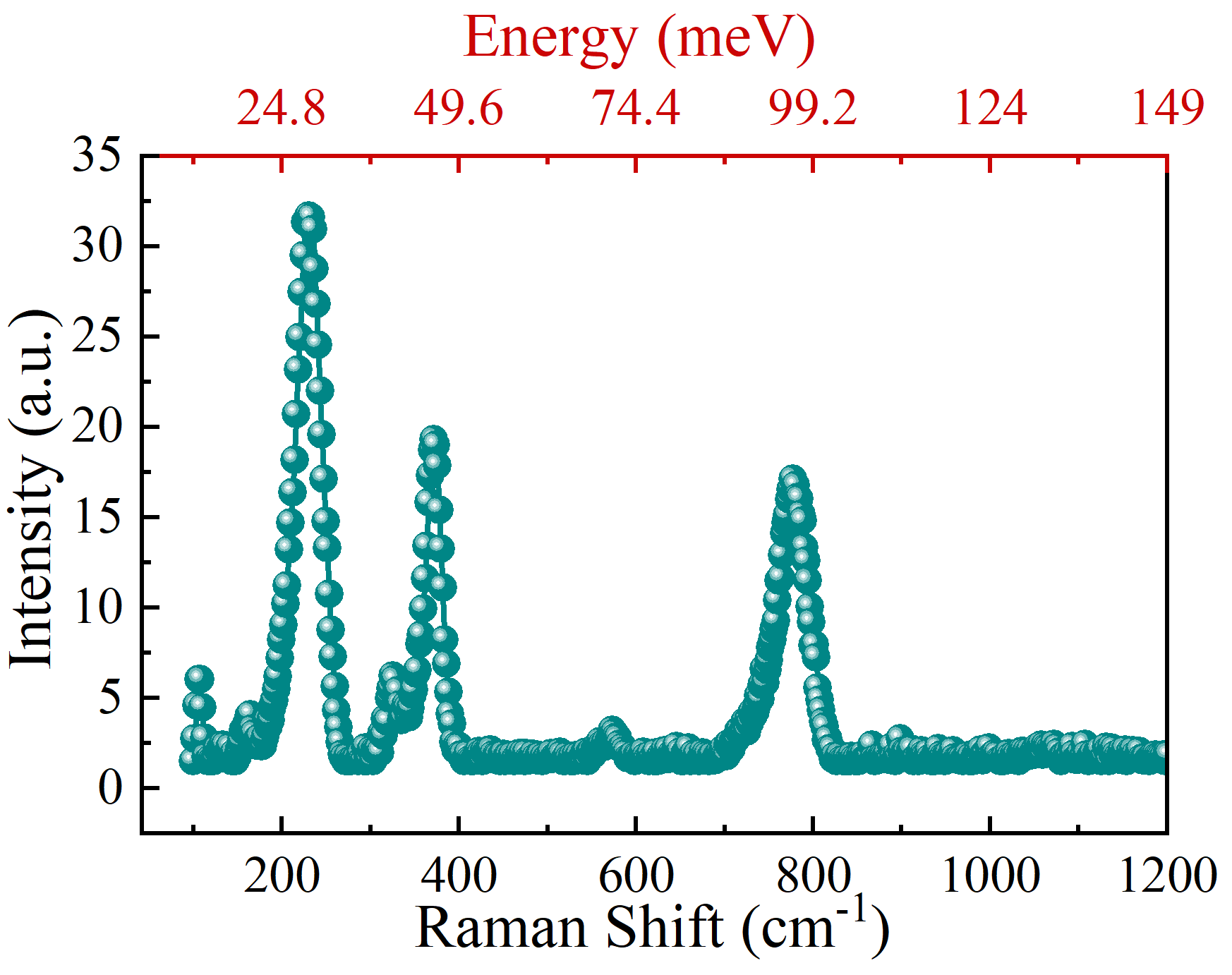}
\caption{\label{fig:raman}Room-temperature Raman spectrum of Ba$_3$HoRu$_2$O$_9$. The top axis gives the corresponding energy in meV.}
\end{figure}

To assign the observed Raman peaks and examine their vibrational character, we calculated the $\Gamma$-point phonon modes using phonopy and the MACE-MP-0 force field~\cite{batatia2025foundation,togo2015first}. The calculated eigenvectors were analysed to identify the atomic contributions to each mode. Table II and Fig. 7 present the calculated modes in the energy range covered by the Raman measurements (13-96~meV), while additional low-energy modes, including those that are Raman-silent or fall below the experimental frequency range, are provided in SI Section D. The calculated Raman-active modes are consistent with the observed peaks. The energy deviations are approximately 13\% for the Ru-dominated modes R2 and R3 and less than 10\% for the remaining modes. Deviations of a similar magnitude have also been reported in lattice-dynamics calculations for isostructural Ba$_3$ZnSb$_2$O$_9$~\cite{glamazda2017comparative}.

\begin{table}[!t]
\caption{\label{tab:raman}Observed Raman modes and the nearest calculated Raman-active $\Gamma$-point modes (in cm$^{-1}$). Ba, Ru, O(1) and O(2) give the share of the kinetic energy (\%) carried by Ba, Ru, the face-sharing O(1) and the corner-sharing O(2) atoms. Values are rounded to whole numbers. Ho does not move in any Raman-active mode. For R6 the $E_{1g}$ (529~cm$^{-1}$) and $E_{2g}$ (531~cm$^{-1}$) modes lie close to the $A_{1g}$ mode.}
\begin{ruledtabular}
\begin{tabular}{lcccrrrr}
Mode & Irrep & Exp. & Calc. & Ba & Ru & O(1) & O(2) \\
\hline
R1 & $E_{2g}$ & 107 & 108 & 22 & 9  & 1  & 68 \\
R2 & $E_{1g}$ & 164 & 186 & 0  & 82 & 2  & 16 \\
R3 & $A_{1g}$ & 231 & 202 & 0  & 82 & 7  & 11 \\
R4 & $E_{1g}$ & 325 & 324 & 0  & 1  & 55 & 43 \\
R5 & $A_{1g}$ & 370 & 335 & 1  & 5  & 2  & 92 \\
R6 & $A_{1g}$ & 574 & 537 & 0  & 9  & 90 & 1  \\
R7 & $A_{1g}$ & 774 & 741 & 0  & 4  & 1  & 96 \\
\end{tabular}
\end{ruledtabular}
\end{table}

The lowest-energy mode, R1, is the only mode in the measured range with a substantial Ba contribution (22\%). In this mode, Ba atoms move within the $ab$ plane against the corner-sharing oxygen atoms, while the polyhedra undergo little internal distortion. The next two modes, R2 and R3 at 20 and 29~meV, are dominated by Ru motion within the Ru$_2$O$_9$ dimer. In R2, the two Ru atoms of a dimer move in antiphase in the $ab$ plane, whereas in R3 they move in opposite directions along the $c$ axis, which changes the short Ru-Ru distance. The intermediate-energy modes R4 and R5 (40-46~meV) are almost purely oxygen modes. R4 is a mixed bending and stretching motion of both oxygen sites, and R5 is mainly an O-Ru-O bending of the corner-sharing oxygen atoms. These modes are sensitive to the Ru-O bond lengths and Ru-O-Ru bond angles, which play a crucial role in the superexchange interactions within the Ru sublattice~\cite{kolev2002raman,iliev1999raman}. The high-energy Raman modes R6 and R7 (71 and 96~meV) are Ru-O stretching vibrations: R6 is the in-plane breathing of the face-sharing oxygen atoms and R7 the stretching of the corner-sharing oxygen atoms, and their high energy reflects the stiffness of the Ru-O bonds. The oxygen arrows in Fig. 7 differ in length and direction because the face-sharing and corner-sharing oxygen sites are not equivalent and because oxygen is much lighter than Ru. Ho sits on an inversion centre and does not move in any Raman-active mode. The same sequence, Ba modes at low energy, $M_2$O$_9$ dimer motion at intermediate energy and oxygen modes at high energy, is found in Ba$_3$ZnSb$_2$O$_9$, which has the same structure~\cite{glamazda2017comparative}.
\begin{figure}[t]
\centering
\includegraphics[width=1\columnwidth]{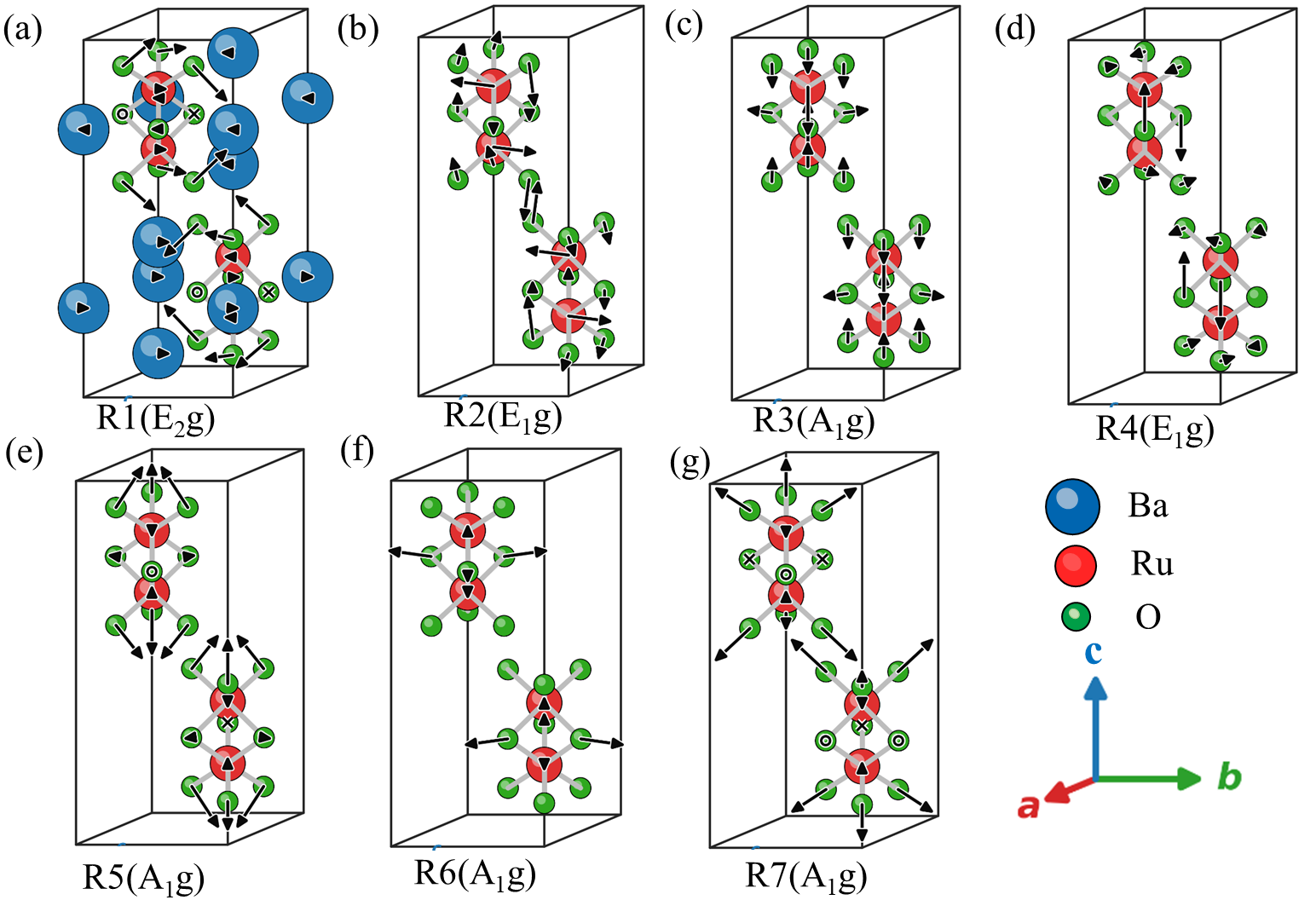}
\caption{\label{fig:eigen}Calculated $\Gamma$-point eigenvectors of the Raman-active modes assigned to the observed lines; (a)-(g) correspond to R1-R7 of Table~\ref{tab:raman}. Arrows show the direction and relative amplitude of the atomic displacements; a dot or a cross marks a displacement toward or away from the viewer.}
\end{figure}

The same calculation also gives the lowest zone-centre optical modes, which lie below the measured Raman range (SI Table S7, Fig. S10). The two lowest, the $E_{2u}$ mode at 5.7~meV and the $A_{2g}$ mode at 6.6~meV, are silent in Raman scattering; the $A_{2g}$ mode is the rigid-unit mode ($\Gamma_2^+$) of the 6H structure~\cite{simpson2025octahedral}, in which the HoO$_6$ and Ru$_2$O$_9$ units rotate about $c$ as rigid bodies. Both modes lie at the upper edge of the low-energy magnetic band (6.2~meV). Phonon intensity increases with $Q$, whereas the intensity of this band decreases (Fig. S1), so these phonons do not account for the excitation below 6.2~meV (Fig. 2). We paid particular attention to the spectral region corresponding to the broad INS excitation centred near 39~meV at low temperature, which shifts to 37.25~meV at 300~K. In this energy range, the Raman spectrum exhibits two well-defined optical phonon modes, R4 (40.3~meV) and R5 (45.9~meV), providing independent experimental evidence for lattice vibrations within the same energy window. R5 is mainly a motion of the corner-sharing O(2) atoms, which also form the HoO$_6$ octahedron. Since Raman spectroscopy probes only Brillouin-zone-centre ($\Gamma$-point) excitations, whereas INS measures momentum-dependent excitations throughout the entire Brillouin zone, the two techniques provide complementary information~\cite{rosalin2024magnon}. The Raman results therefore confirm that optical phonons contribute to the INS intensity near 39~meV. They cannot, however, account for the magnetic part of this feature, which is strongest at low $|Q|$ and follows the Ho$^{3+}$ form factor (Fig. S6(e), and which we attribute to the Ho$^{3+}$ CEF transitions (Figs. 4 and5).

\subsection{Machine-learned phonon calculations}

We performed phonon calculations using a machine-learned force field (MLFF)~\cite{yang2024mattersim} within the INSPIRED framework~\cite{cheng2023direct} to distinguish the lattice contribution to the high-energy excitations observed in the INS measurements. This methodology allows accurate modelling of lattice vibrations over the full Brillouin zone while retaining the accuracy of first-principles calculations (e.g., \textit{ab initio} DFT), but at a much lower computational cost~\cite{cheng2023direct}. This is an advantage for Ba$_3$HoRu$_2$O$_9$, whose 30-atom unit cell contains correlated Ru $4d$ and Ho $4f$ electrons; a first-principles phonon calculation would require DFT+$U$ with an ordered magnetic configuration in large supercells. We used the experimentally determined low- and room-temperature crystal structures as input for structural optimization to minimize the residual forces. The resulting optimized structure exhibits no imaginary phonon frequencies, confirming its dynamical stability. The calculated phonon dispersion relations are shown in Fig. 8(a), with the phonon energies extending up to approximately 100~meV.
\begin{figure}[t]
\centering
\includegraphics[width=0.8\columnwidth]{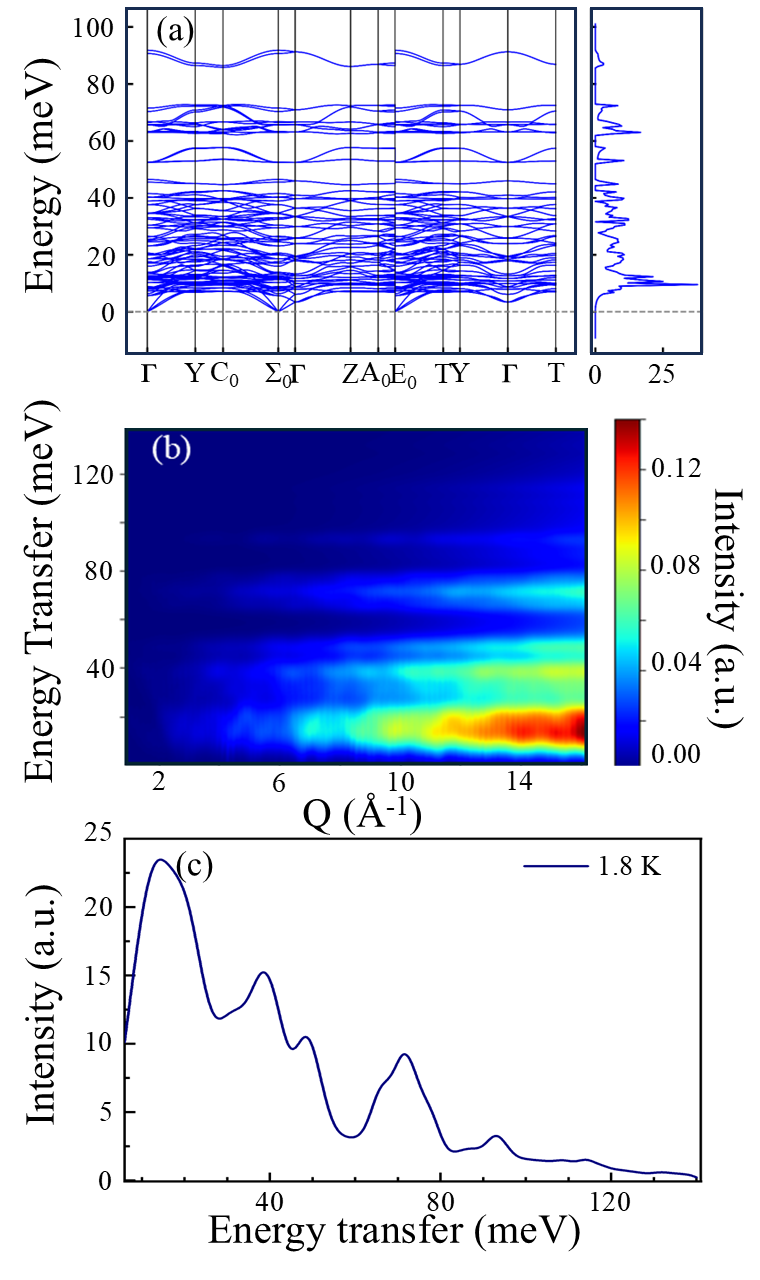}
\caption{\label{fig:mlff}(a) Phonon dispersion of Ba$_3$HoRu$_2$O$_9$ calculated with the machine-learned force field along high-symmetry directions of the Brillouin zone (left), and the corresponding phonon density of states (right). (b) Calculated powder-averaged phonon scattering intensity at 1.8~K. (c) Calculated $|Q|$-integrated intensity at 1.8~K for $8 < |Q| < 16$~\AA$^{-1}$ as a function of energy transfer.}
\end{figure}

Using the MLFF, we computed the powder-averaged dynamical structure factor, including one-phonon and multiphonon contributions, and convolved it with the instrumental energy and momentum resolution to facilitate direct comparison with the INS data. The simulated intensity at $T = 1.8$~K is presented in Fig. 8(b), while the corresponding $|Q|$-integrated spectrum over the range $8 < |Q| < 16$~\AA$^{-1}$ is shown in Fig. 8(c)~\cite{cheng2023direct}. The phonon calculations reproduce the phonon excitations observed experimentally at approximately 20, 40, 46, 70-75 and 90-95~meV, in good agreement with the experimental INS spectra for $8 < |Q| < 16$~\AA$^{-1}$, where the neutron response is expected to be dominated by nuclear (phonon) scattering. This agreement confirms that the high-$|Q|$ excitations originate predominantly from lattice vibrations.  At low $|Q|$, by contrast, the calculated phonon intensity is weak, whereas the measured intensity near 39~meV is largest there. This difference shows that, in addition to phonons, another excitation contributes to the neutron response in this energy range. It agrees with the momentum dependence of the 39~meV feature, whose magnetic part follows the Ho$^{3+}$ form factor (Fig. S6(e)), and with the CEF calculations, which place the strongest Ho$^{3+}$ transitions between 39 and 45~meV (Figs. 4 and 5). Consequently, the broad excitation observed near 39~meV is interpreted as arising from overlapping phonon and Ho$^{3+}$ crystal-field excitations, while the high-$|Q|$ ($8 < |Q| < 16$~\AA$^{-1}$) response is predominantly phononic.

\section{Conclusion}
We investigated the elementary excitations of the hexagonal 6H-perovskite multiferroic Ba$_3$HoRu$_2$O$_9$, a spin-driven ferroelectric system, using a combination of inelastic neutron scattering, linear spin-wave theory, crystal-electric-field calculations, Raman spectroscopy, and machine-learned phonon calculations. The complementary experimental and theoretical approaches provide a comprehensive understanding of the magnetic and lattice dynamics over a broad energy range. The low-energy excitation below 6.2~meV, observed only in the low-temperature magnetic phase, is reproduced by linear spin-wave calculations in which the intradimer Ru-Ru exchange ($J_1 = 1.57$~meV) and the Ru-Ho exchange ($J_2 = 1.45$~meV) are of comparable size and the interdimer terms are small. It is therefore attributed to collective spin waves of the coupled Ru-Ho magnetic network rather than to excitations of isolated Ru$_2$O$_9$ dimers. In contrast, several broad excitations centred near 20, 39, 70, and 90~meV persist well above the magnetic ordering temperatures, demonstrating that they cannot be explained by conventional spin-wave excitations. The magnetic part of the 39~meV feature, isolated by subtracting the Bose-corrected 300~K data, follows the Ho$^{3+}$ magnetic form factor. Crystal-electric-field calculations based on the Stevens formalism show that Ho$^{3+}$ supports a group of strong crystal-field transitions in the same energy range. The HoO$_6$ octahedron is almost regular, and the crystal-field Hamiltonian is a cubic field plus a small trigonal term to within 1\%. This explains why the calculated transition intensities differ by orders of magnitude and why only the strongest group of transitions is expected to be observed. A uniform scaling of the point-charge parameters places this group at 39-45~meV, and its spread accounts for much of the measured width. Raman spectroscopy and machine-learned phonon calculations show that the features at 20, 70 and 90~meV, and the high-$|Q|$ part of the 39~meV feature, arise from lattice vibrations. The high-energy neutron response of Ba$_3$HoRu$_2$O$_9$ is therefore most reasonably interpreted as overlapping Ho$^{3+}$ crystal-field excitations and lattice vibrations. Ru$_2$O$_9$ dimer excitations, which are intrinsic to this family, may contribute only weakly.

Because these crystal-field and lattice excitations overlap in energy, they may also be coupled. Such a coupling requires that the phonon modulates the crystal field and has a symmetry compatible with the crystal-field transition. The crystal-field transitions within the $J = 8$ multiplet are even under inversion at the Ho site, so only even ($g$) phonons can couple to them. The Raman-active modes at 40.3~meV ($E_{1g}$) and 45.9~meV ($A_{1g}$) are $g$ modes in this range, and the $A_{1g}$ mode mainly moves the oxygen atoms that form the HoO$_6$ octahedron, so a magnetoelastic coupling is allowed by symmetry. The present powder INS and room-temperature Raman data cannot establish whether the modes actually mix. Temperature-dependent Raman measurements across the magnetic transitions, or INS on single crystals, would be needed to test this. The present work provides a microscopic framework for understanding the complex excitation spectrum of rare-earth ruthenates and establishes Ba$_3$HoRu$_2$O$_9$ as an important model system for investigating the interplay between localized $4f$ crystal-field states, molecular magnetism associated with transition-metal dimers, and lattice dynamics.

\begin{acknowledgments}
T.B. acknowledges financial support from the Science and Engineering Research Board (SERB), Anusandhan National Research Foundation (ANRF), Government of India, under Project No.~SRG/2022/000044. T.B. also acknowledges the Science and Technology Facilities Council (STFC, UK) for neutron beam time on the MARI instrument (Proposal No.~RB2520463) at the ISIS Neutron and Muon Source, and the computing facilities at Oak Ridge National Laboratory (ORNL) for the machine-learning-based phonon calculations. Spin-wave calculations using SpinW were performed with support from the MATLAB Academic License available at Rajiv Gandhi Institute of Petroleum Technology (RGIPT). D.T.A. acknowledges financial support from the Engineering and Physical Sciences Research Council (EPSRC, UK) under Grant No.~EP/W00562X/1. J.S. acknowledges the Ramanujan Fellowship from the Science and Engineering Research Board (SERB), Department of Science and Technology (DST), India (Grant No.~RJF/2019/000046).
\end{acknowledgments}
\section*{Competing interests}
The authors declare no competing financial interests.

\bibliography{refs}
\clearpage

\setcounter{figure}{0}
\setcounter{table}{0}

\renewcommand{\thefigure}{S\arabic{figure}}
\renewcommand{\thetable}{S\arabic{table}}

\section{Spin-wave simulations}

\begin{figure}[h]
\includegraphics[width=\columnwidth]{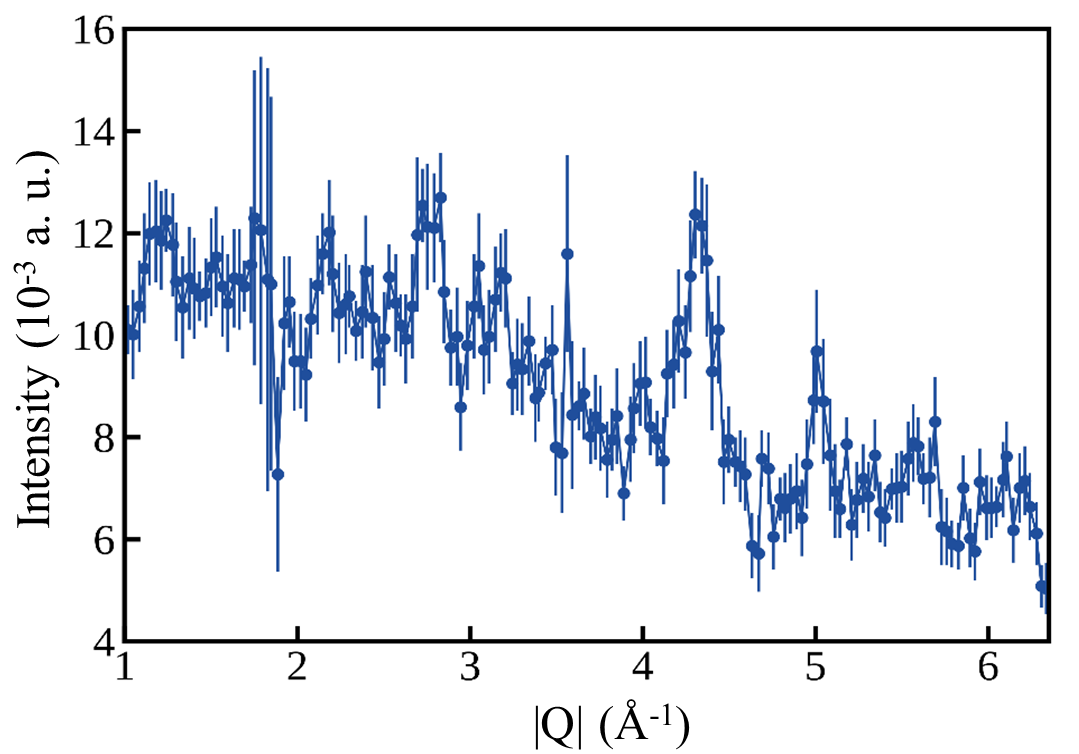}
\caption{\label{fig:S1}$|Q|$ dependence of the intensity integrated between 2 and 7~meV at 1.8~K. The intensity decreases by about a factor of two between 1.2 and 6~\AA$^{-1}$ and does not increase at high $|Q|$, whereas phonon scattering would increase with $|Q|$.}
\end{figure}
\begin{figure}[h]
\includegraphics[width=\columnwidth]{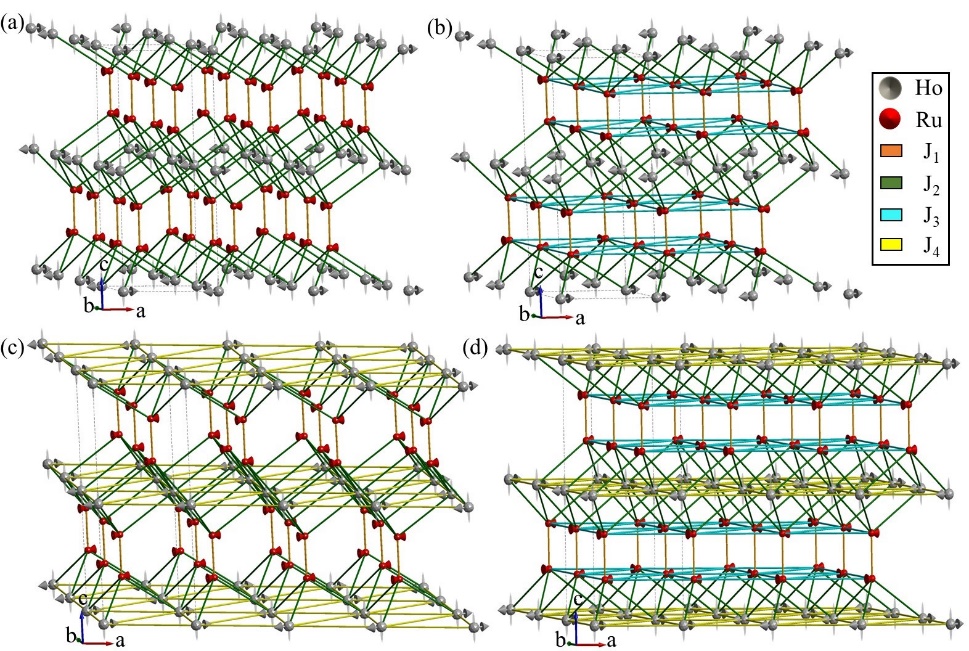}
\caption{\label{fig:S2}Magnetic structure of Ba$_3$HoRu$_2$O$_9$ obtained from SpinW for different sets of exchange interactions. The single-ion anisotropy $D_{zz} = -0.25$~meV is included in every case. (a) Only $J_1$ and $J_2$. (b) $J_1$, $J_2$ and $J_3$, with $J_4 = 0$. (c) $J_1$, $J_2$ and $J_4$, with $J_3 = 0$. (d) All four interactions. Only the model with all four exchange interactions (d) reproduces the magnetic structure found by neutron diffraction [Ref.~16 of the main text].}
\end{figure}
\begin{figure}[h]
\includegraphics[width=\columnwidth]{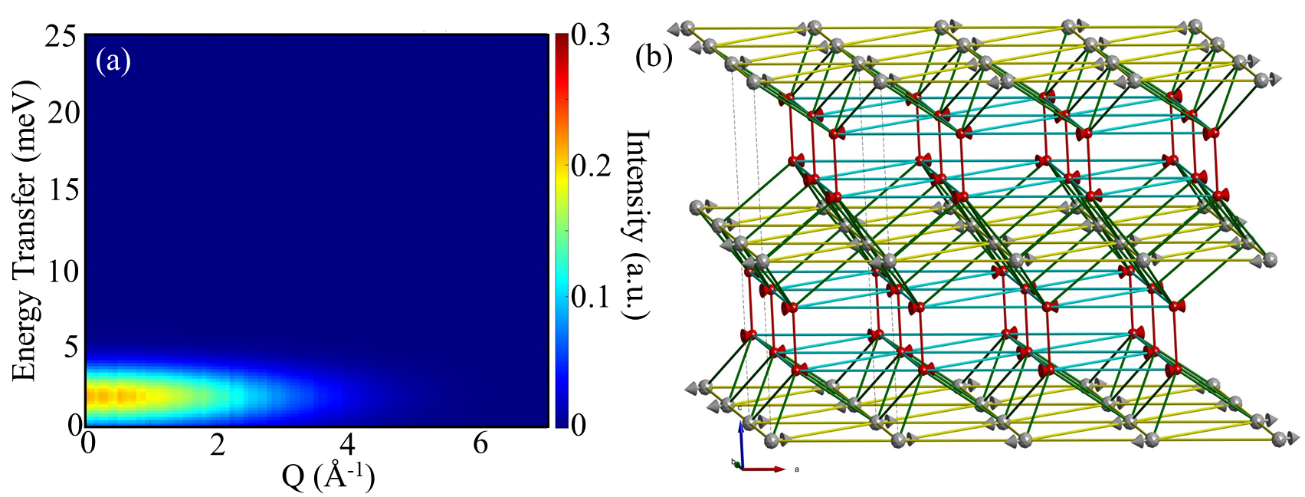}
\caption{\label{fig:S3}(a) Calculated spin-wave spectrum with the single-ion anisotropy set to zero ($D_{zz} = 0$). (b) Magnetic structure obtained from SpinW for the same exchange interactions ($J_1$-$J_4$), also with $D_{zz} = 0$.}
\end{figure}
\begin{figure}[h]
\includegraphics[width=\columnwidth]{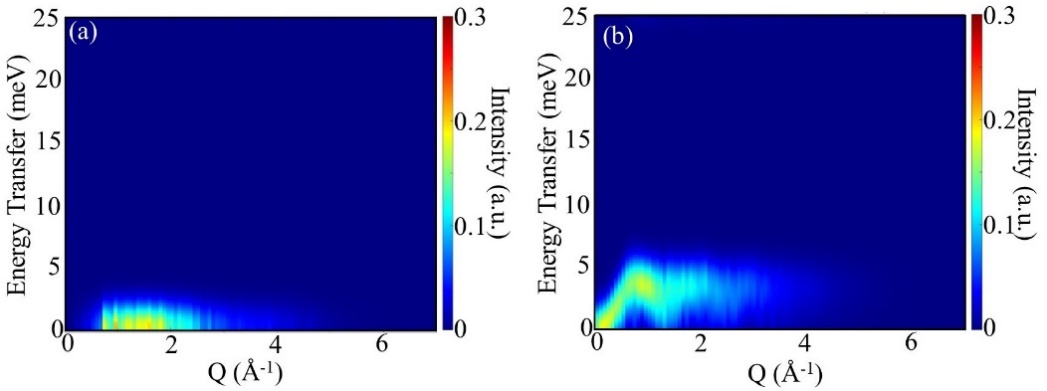}
\caption{\label{fig:S4}Calculated spin-wave spectrum with (a) only the intradimer Ru-Ru exchange $J_1$ and (b) only the Ru-Ho exchange $J_2$. $J_1$ alone gives weak intensity at low energy, and $J_2$ alone gives a dispersive mode reaching about 5~meV.}
\end{figure}

\section{High-energy excitations}

\begin{figure}[h]
\includegraphics[width=\columnwidth]{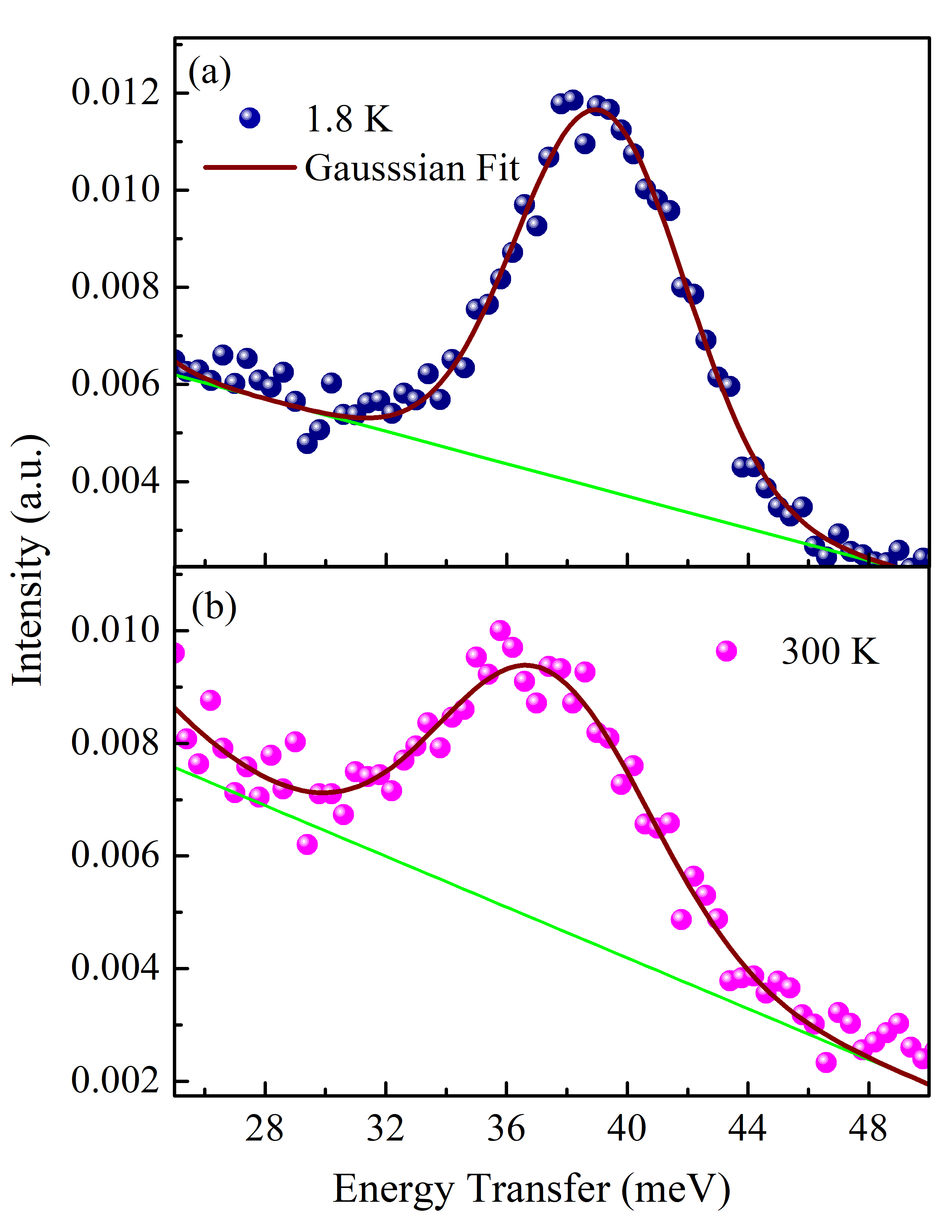}
\caption{\label{fig:S5}Gaussian fits of the momentum-integrated INS spectra ($1 \le |Q| \le 8$~\AA$^{-1}$, $E_i = 160$~meV) at (a) 1.8~K and (b) 300~K. Solid line: Gaussian plus linear background; green line: background.}
\end{figure}

\begin{figure}[h]
\includegraphics[width=\columnwidth]{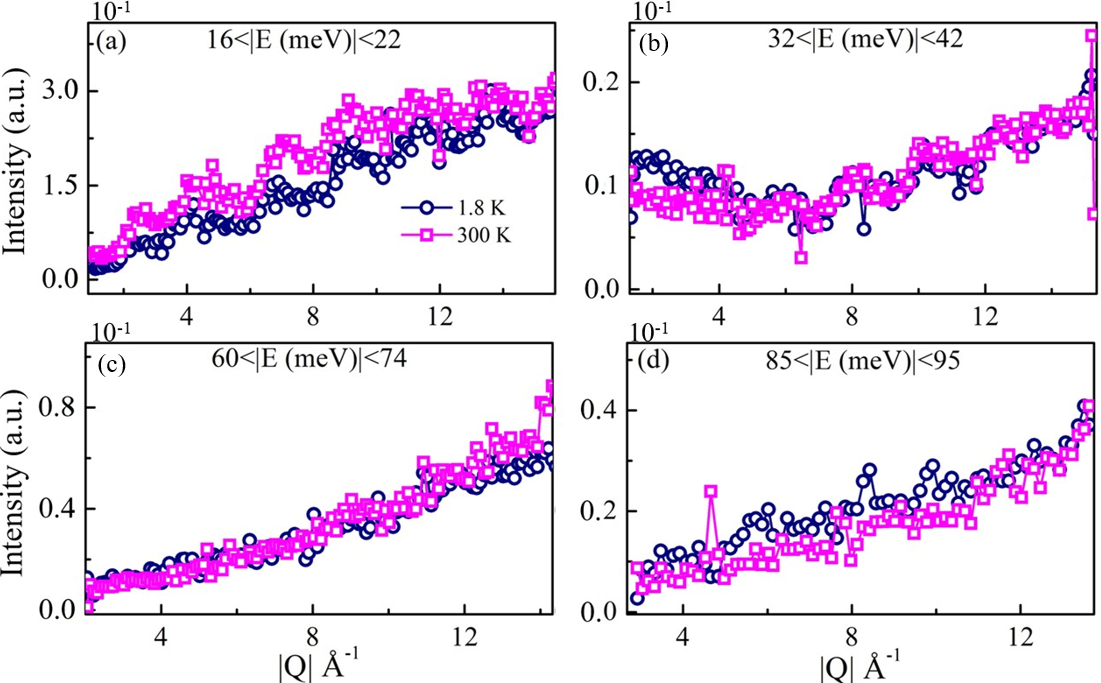}\\[2pt]
\includegraphics[width=0.55\textwidth]{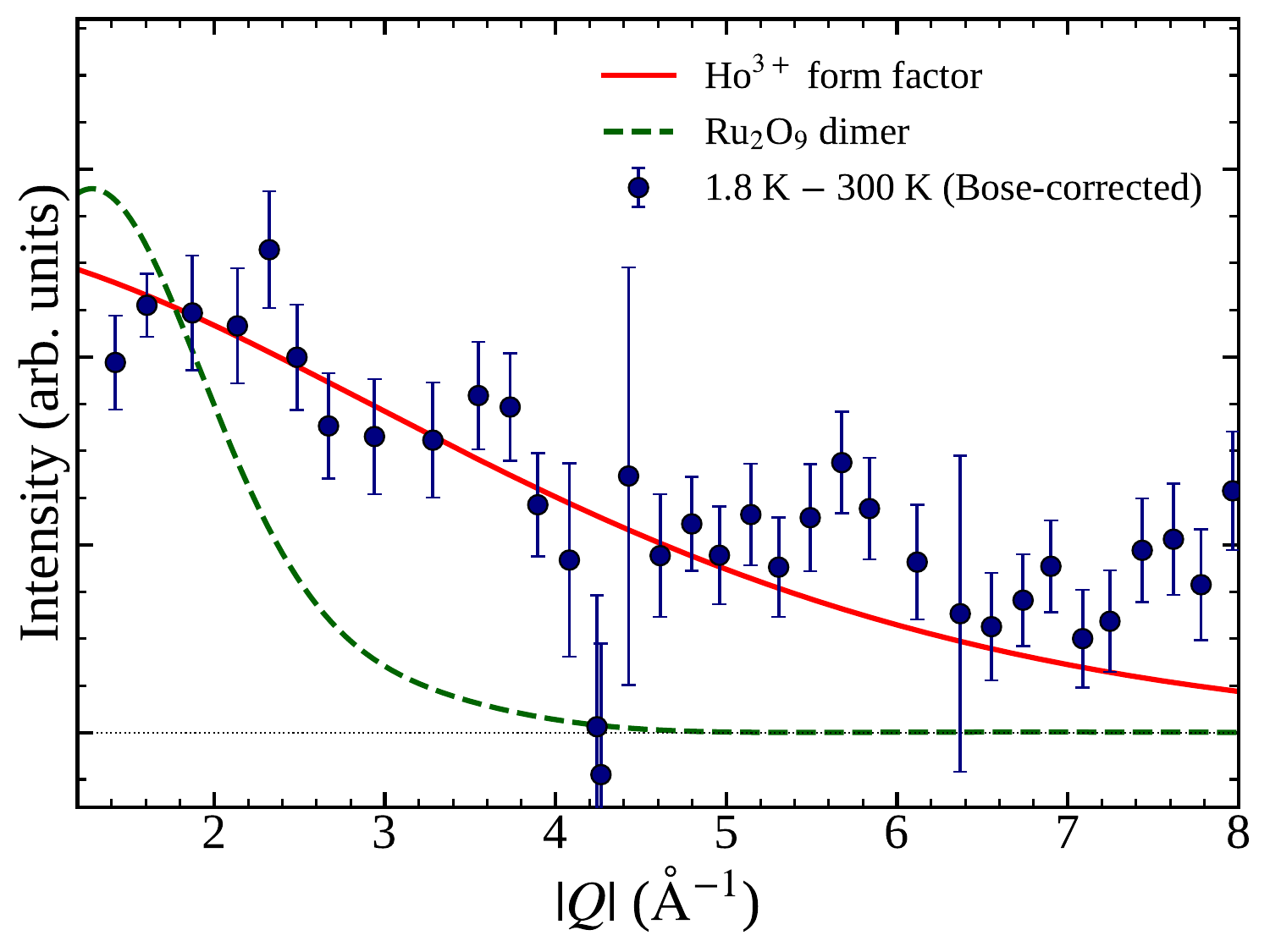}
\caption{\label{fig:S6}(a)-(d) $|Q|$ dependence of the INS intensity at 1.8~K and 300~K, integrated over (a) 16-22~meV, (b) 32-42~meV, (c) 60-74~meV and (d) 85-95~meV. (e) Magnetic part of the 39~meV excitation, obtained as the difference between the 1.8~K and the Bose-corrected 300~K data (integrated over 32-42~meV), compared with fits using the Ho$^{3+}$ magnetic form factor (solid line) and the $Q$ dependence expected for a Ru$_2$O$_9$ dimer excitation (dashed line). Fits were made between 1.4 and 5.5~\AA$^{-1}$.}
\end{figure}


\subsection*{Separation of the magnetic and phonon contributions to the 39~meV feature}

The measured intensity in the 32-42~meV window is the sum of a phonon and a magnetic part,
\begin{equation}
I(Q,T) = I_{\mathrm{ph}}(Q,T) + I_{\mathrm{mag}}(Q,T).
\end{equation}
The phonon intensity on the neutron energy-loss side scales with the Bose factor,
\begin{equation}
I_{\mathrm{ph}}(Q,T) = [n(\omega,T)+1]\, I_{\mathrm{ph}}(Q,0), \qquad n(\omega,T) = \frac{1}{e^{\hbar\omega/k_BT}-1}.
\end{equation}
At 1.8~K, $n \approx 0$ for all energies of interest, while at 300~K ($k_BT = 25.85$~meV) and $\hbar\omega = 39$~meV, $n+1 = 1.284$ (1.41 and 1.25 at the edges of the window, 32 and 42~meV). The difference
\begin{equation}
\Delta I(Q) = I(Q,1.8\,\mathrm{K}) - \frac{I(Q,300\,\mathrm{K})}{n(\omega,300\,\mathrm{K})+1}
\end{equation}
therefore removes the phonon part, and leaves
\begin{equation}
\Delta I(Q) = I_{\mathrm{mag}}(Q,1.8\,\mathrm{K}) \left[ 1 - \frac{r}{n+1} \right], \qquad r = \frac{I_{\mathrm{mag}}(300\,\mathrm{K})}{I_{\mathrm{mag}}(1.8\,\mathrm{K})}.
\end{equation}
For the Ho$^{3+}$ crystal-field transitions, $r$ follows from the thermal population of the levels in the scaled CEF model (Table S6). Summing all transitions between 32 and 42~meV gives $r = 0.67$, so that
\begin{equation}
1 - \frac{r}{n+1} = 1 - \frac{0.67}{1.284} \approx 0.48 .
\end{equation}
About half of the magnetic intensity thus remains in $\Delta I(Q)$. Since this factor does not depend on $Q$, $\Delta I(Q)$ has the same $Q$ dependence as the magnetic scattering. The subtraction is reliable up to about 5.5~\AA$^{-1}$; at higher $Q$ the Debye-Waller factor and multiphonon scattering at 300~K make it incomplete.

The magnetic part was compared with two forms. For Ho$^{3+}$ we used the dipole approximation,
\begin{equation}
f_{\mathrm{Ho}}(Q) = \langle j_0 \rangle + \frac{2-g_J}{g_J}\langle j_2 \rangle, \qquad g_J = \tfrac{5}{4},
\end{equation}
with the radial integrals $\langle j_0\rangle$ and $\langle j_2\rangle$ of Brown [International Tables for Crystallography, Vol.~C, Sec.~4.4.5]. A transition of a Ru$_2$O$_9$ dimer that changes its total spin gives~[Furrer and G\"udel, J. Magn. Magn. Mater. \textbf{14}, 256 (1979); Furrer and Waldmann, Rev. Mod. Phys. \textbf{85}, 367 (2013)]
\begin{equation}
I_{\mathrm{dimer}}(Q) \propto |f_{\mathrm{Ru}}(Q)|^2 \left[ 1 - \frac{\sin(Qd)}{Qd} \right], \qquad d = 2.55~\text{\AA},
\end{equation}
which vanishes at $Q \to 0$ and peaks near 1.3~\AA$^{-1}$. The fraction of the magnetic intensity that remains at higher $Q$, relative to its value at $Q_0 = 1.7$~\AA$^{-1}$, is compared below with both forms.

\begin{center}
\begin{tabular}{cccc}
\hline\hline
$|Q|$ (\AA$^{-1}$) & Measured $\Delta I$ (\%) & Ho$^{3+}$ $|f(Q)|^2$ (\%) & Ru$_2$O$_9$ dimer (\%) \\
\hline
1.7 & 100 & 100 & 100 \\
2   & $97 \pm 10$ & 95 & 75 \\
3   & $69 \pm 10$ & 75 & 15 \\
4   & $53 \pm 8$  & 55 & 3 \\
5   & $47 \pm 7$  & 38 & $<1$ \\
\hline\hline
\end{tabular}
\end{center}

The measured fraction follows the Ho$^{3+}$ form factor within the uncertainty up to 4~\AA$^{-1}$, while the dimer form has already fallen to 15\% at 3~\AA$^{-1}$. At 5~\AA$^{-1}$ the measured value lies slightly above the Ho$^{3+}$ curve, consistent with the onset of incomplete phonon subtraction. Fits of the data between 1.4 and 5.5~\AA$^{-1}$ with a single free scale factor give a reduced $\chi^2$ of 1.0 for the Ho$^{3+}$ form factor and 13.4 for the dimer form [Fig. S6(e)]; a fit with both components gives no dimer contribution within the uncertainty.

\section{Crystal-electric-field analysis of Ho$^{3+}$ excitations}

\subsection{Initial point-charge parameters}

\begin{table}[h]
\caption{\label{tab:S1}Initial Stevens crystal-field parameters for Ho$^{3+}$ in Ba$_3$HoRu$_2$O$_9$, obtained from the point-charge calculation using the experimentally refined crystal structure.}
\begin{ruledtabular}
\begin{tabular}{lc}
Parameter & Value (meV) \\
\hline
$B_2^0$ & $8.367\times10^{-2}$ \\
$B_4^0$ & $6.09\times10^{-4}$ \\
$B_4^3$ & $-1.747\times10^{-2}$ \\
$B_6^0$ & $-1.39\times10^{-6}$ \\
$B_6^3$ & $-1.852\times10^{-5}$ \\
$B_6^6$ & $-1.383\times10^{-5}$ \\
\end{tabular}
\end{ruledtabular}
\end{table}

\begin{table}[h]
\caption{\label{tab:S2}All 11 crystal-field levels (17 states) of the Ho$^{3+}$ multiplet for the initial parameters (Table S1). Energies are measured from the ground-state doublet. The relative INS intensity is the strength of the transition from the ground doublet to the level, summed over the degenerate states. The first row is the transition within the ground doublet; the first four levels lie within 1~meV, inside the elastic line.}
\begin{ruledtabular}
\begin{tabular}{cccc}
Level & Degeneracy & Energy (meV) & Intensity \\
\hline
0 & 2 & 0 & 0.29 \\
1 & 2 & 0.122 & 12.1 \\
2 & 1 & 0.235 & 6.73 \\
3 & 1 & 0.470 & 0.126 \\
4 & 1 & 25.815 & 0.823 \\
5 & 2 & 26.876 & 0.849 \\
6 & 2 & 28.807 & 0.498 \\
7 & 1 & 29.533 & 0.189 \\
8 & 1 & 35.512 & 0.033 \\
9 & 2 & 38.042 & 0.074 \\
10 & 2 & 46.765 & 0.016 \\
\end{tabular}
\end{ruledtabular}
\end{table}

\begin{figure}[h]
\includegraphics[width=\columnwidth]{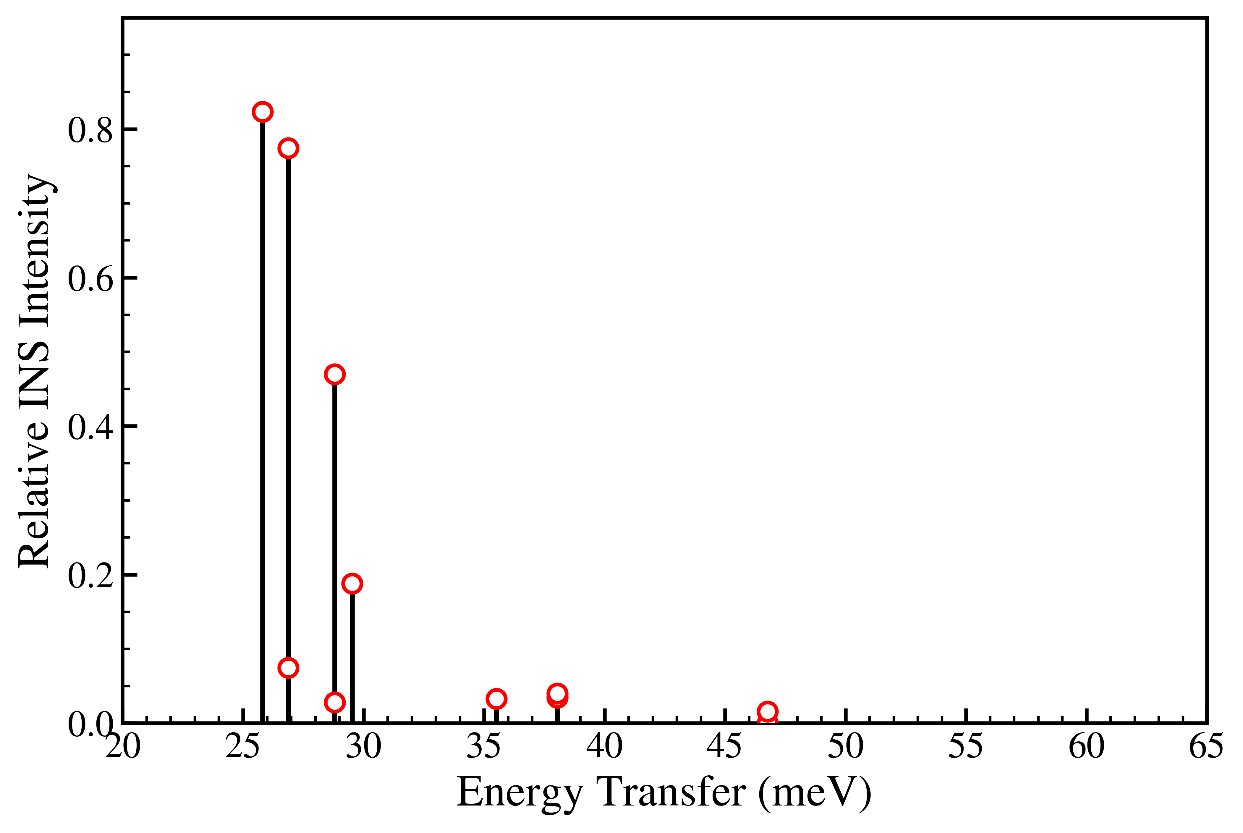}
\caption{\label{fig:S7}Calculated relative INS transition intensities for Ho$^{3+}$ from the initial point-charge CEF model. Vertical black lines mark the calculated transition energies from the ground-state doublet, and open red circles the corresponding relative intensities. The strongest calculated transitions lie near 25-30~meV, followed by weaker excitations at higher energies.}
\end{figure}

\subsection{Comparison with a cubic crystal field}

\begin{table}[h]
\caption{\label{tab:S3}Comparison of the $D_{3d}$ crystal-field Hamiltonian with a cubic field (threefold axis along $c$) plus a trigonal $B_2^0$ term. The deviation is the norm of the difference between the two Hamiltonians relative to the norm of the full Hamiltonian.}
\begin{ruledtabular}
\begin{tabular}{lc}
Parameter set & Deviation from cubic + $B_2^0$ \\
\hline
Point-charge (this work) & 0.8\% \\
Random $D_{3d}$ parameters & 31-98\% \\
\end{tabular}
\end{ruledtabular}
\end{table}

\subsection{Scaling analysis}

\begin{table}[h]
\caption{\label{tab:S4}Stevens parameters after uniform scaling by 1.51.}
\begin{ruledtabular}
\begin{tabular}{lc}
Parameter & Scaled value (meV) \\
\hline
$B_2^0$ & $1.263\times10^{-1}$ \\
$B_4^0$ & $9.20\times10^{-4}$ \\
$B_4^3$ & $-2.64\times10^{-2}$ \\
$B_6^0$ & $-2.10\times10^{-6}$ \\
$B_6^3$ & $-2.80\times10^{-5}$ \\
$B_6^6$ & $-2.09\times10^{-5}$ \\
\end{tabular}
\end{ruledtabular}
\end{table}

\begin{table}[h]
\caption{\label{tab:S5}Energy and relative INS intensity of the lowest strong Ho$^{3+}$ CEF transition as a function of the uniform scaling factor applied to the Stevens parameters. The transition shifts linearly to higher energy, while its intensity remains nearly unchanged. The factor 1.51 places it at the observed INS peak near 39~meV.}
\begin{ruledtabular}
\begin{tabular}{ccc}
Scaling factor & Energy (meV) & Intensity \\
\hline
1.20 & 30.978 & 0.8816 \\
1.25 & 32.269 & 0.8953 \\
1.30 & 33.560 & 0.9087 \\
1.35 & 34.850 & 0.9218 \\
1.40 & 36.141 & 0.9346 \\
1.45 & 37.432 & 0.9471 \\
1.50 & 38.723 & 0.9593 \\
1.55 & 40.013 & 0.9712 \\
1.60 & 41.304 & 0.9828 \\
1.65 & 42.595 & 0.9940 \\
1.70 & 43.886 & 1.0051 \\
1.75 & 45.176 & 1.0158 \\
1.80 & 46.467 & 1.0262 \\
\end{tabular}
\end{ruledtabular}
\end{table}

\begin{figure}[h]
\includegraphics[width=\columnwidth]{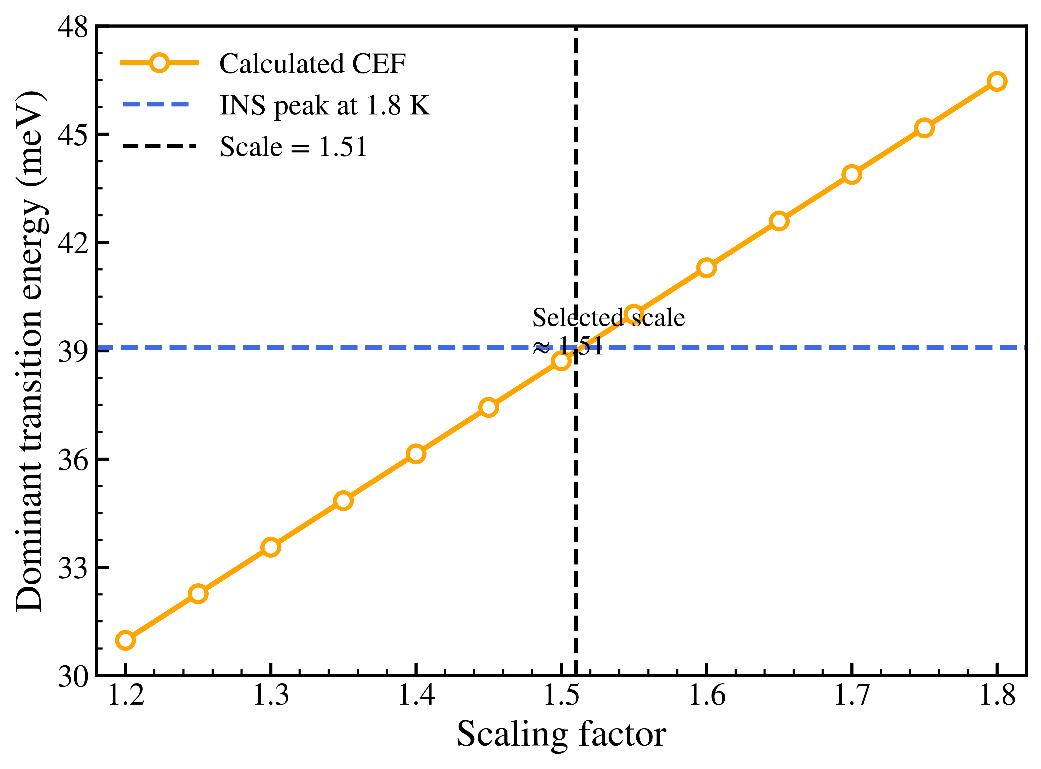}
\caption{\label{fig:S8}Energy of the lowest strong Ho$^{3+}$ CEF transition as a function of the uniform scaling factor applied to the Stevens parameters (orange circles). The horizontal blue dashed line marks the INS peak at 1.8~K, and the vertical black dashed line the selected scaling factor of 1.51. A common factor changes only the overall energy scale and leaves the symmetry of the CEF Hamiltonian unchanged.}
\end{figure}

\begin{table}[h]
\caption{\label{tab:S6}Crystal-field levels after scaling by 1.51, with their relative INS intensities.}
\begin{ruledtabular}
\begin{tabular}{cccc}
Level & Degeneracy & Energy (meV) & Intensity \\
\hline
0 & 2 & 0 & 0.33 \\
1 & 2 & 0.184 & 14.1 \\
2 & 1 & 0.354 & 7.84 \\
3 & 1 & 0.711 & 0.147 \\
4 & 1 & 38.980 & 0.962 \\
5 & 2 & 40.583 & 0.992 \\
6 & 2 & 43.498 & 0.582 \\
7 & 1 & 44.595 & 0.219 \\
8 & 1 & 53.623 & 0.038 \\
9 & 2 & 57.444 & 0.0865 \\
10 & 2 & 70.616 & 0.018 \\
\end{tabular}
\end{ruledtabular}
\end{table}

\begin{figure}[h]
\includegraphics[width=\columnwidth]{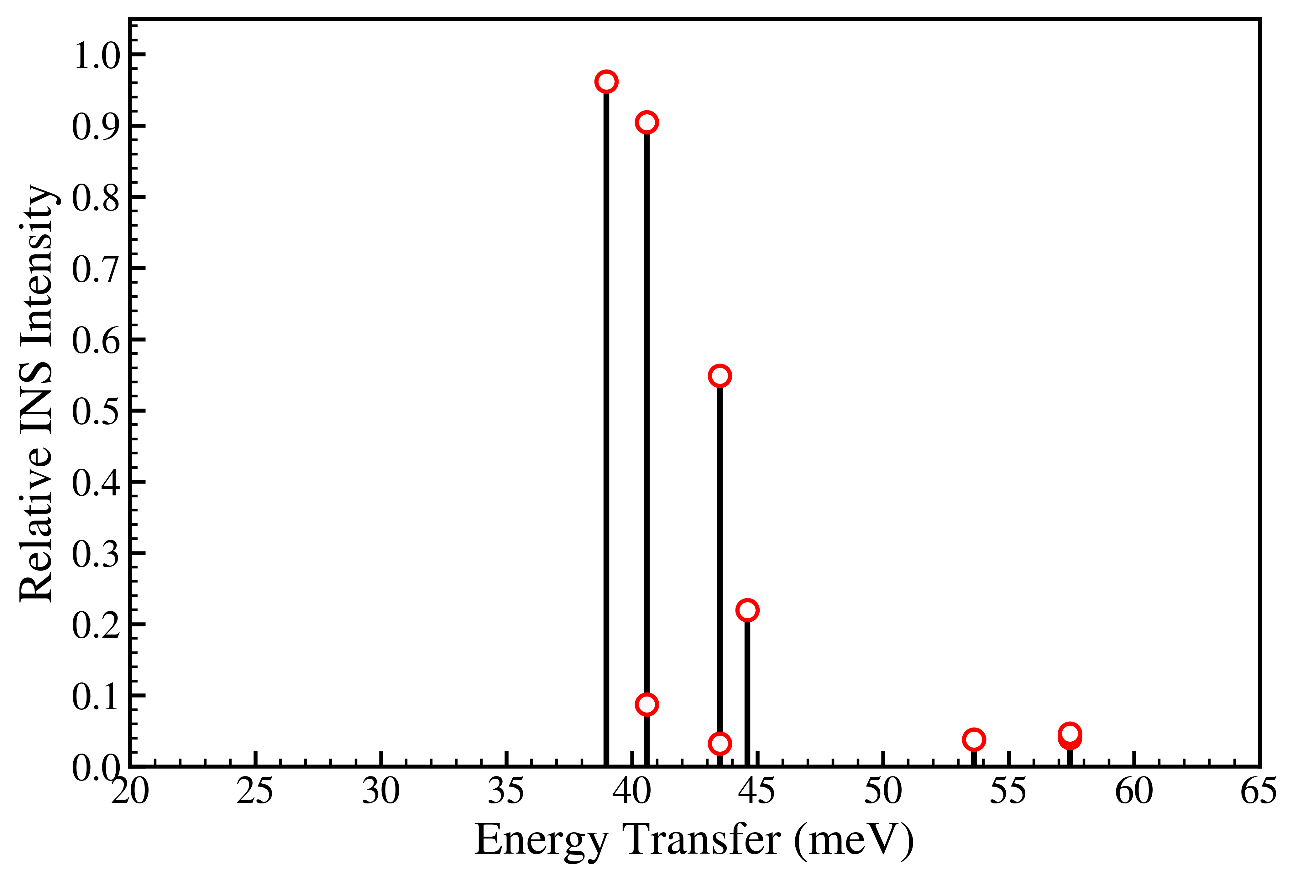}
\caption{\label{fig:S9}Calculated relative INS transition intensities for Ho$^{3+}$ after uniform scaling of the Stevens parameters. Vertical black lines mark the transitions from the ground-state doublet, and open red circles the relative intensities. The strongest transitions lie near 39-41~meV, followed by weaker ones at 43-45~meV and only weak transitions above 50~meV.}
\end{figure}

\section{Lowest zone-centre phonon modes}

Table S7 lists the four lowest $\Gamma$-point optical modes of the phonopy/MACE-MP-0 calculation. They are of interest because they lie at or just above the upper edge of the spin-wave band (6.2~meV). Two of them, L1 ($E_{2u}$) and L2 ($A_{2g}$), are silent in both Raman and infrared scattering and can be observed only by inelastic neutron scattering. L3 ($E_{2g}$) and L4 ($A_{1g}$) are Raman active but lie below the lowest observed Raman line (107~cm$^{-1}$, 13.3~meV). Their eigenvectors are shown in Fig. S10.

\begin{table}[h]
\caption{\label{tab:S7}Lowest $\Gamma$-point optical modes of the calculated phonon spectrum. L2 ($A_{2g}$, $\Gamma_2^+$) is the rigid-unit mode of the 6H structure; L1 and L2 are silent in Raman scattering.}
\begin{ruledtabular}
\begin{tabular}{lcccc}
Mode & Symmetry & cm$^{-1}$ & meV & Raman active \\
\hline
L1 & $E_{2u}$ & 46 & 5.7 & no \\
L2 & $A_{2g}$ ($\Gamma_2^+$) & 53 & 6.6 & no \\
L3 & $E_{2g}$ & 59 & 7.3 & yes \\
L4 & $A_{1g}$ & 71 & 8.8 & yes \\
\end{tabular}
\end{ruledtabular}
\end{table}

\begin{figure}[h]
\includegraphics[width=\columnwidth]{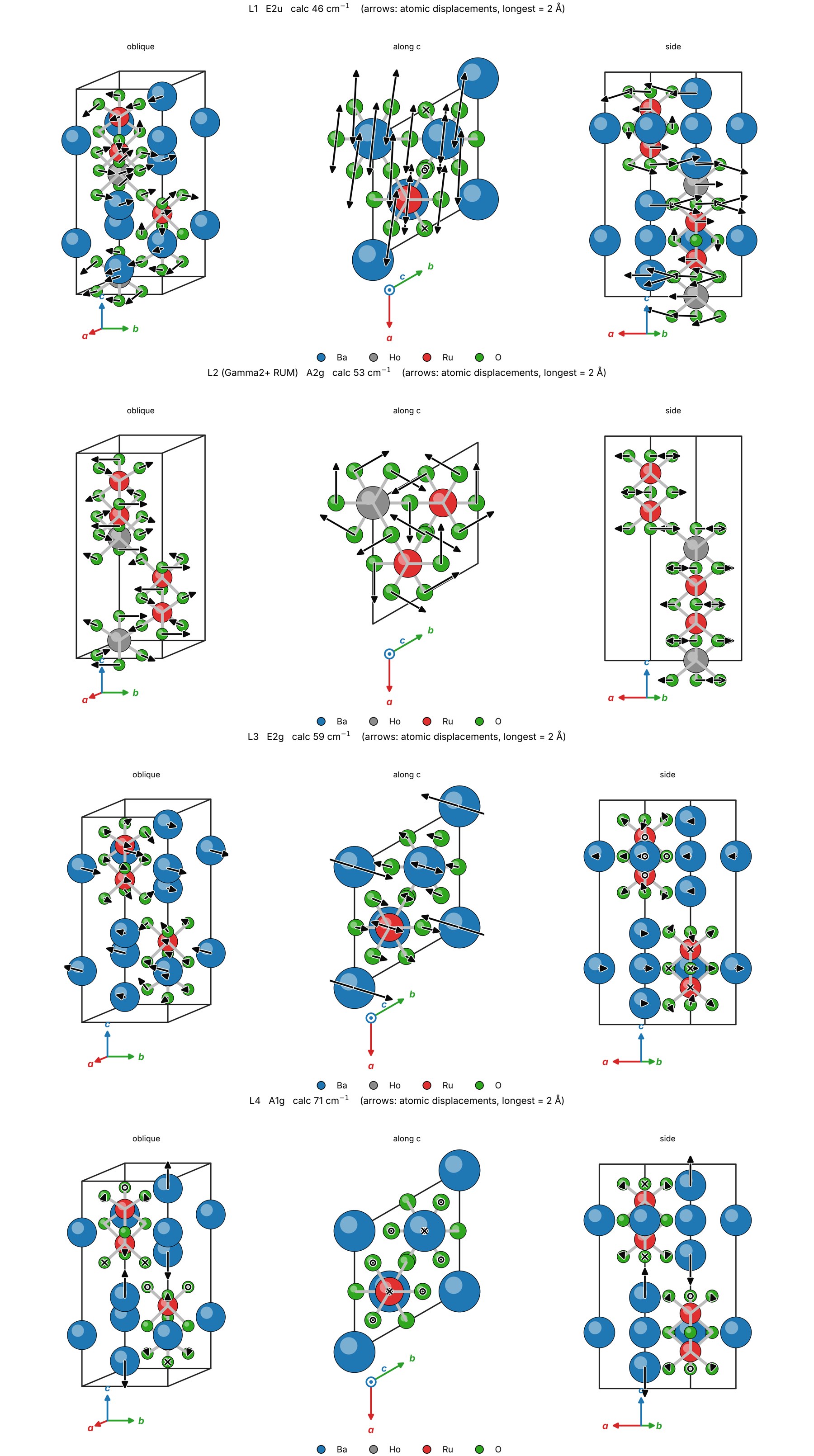}
\caption{\label{fig:S10}Calculated eigenvectors of the four lowest zone-centre modes: (L1) $E_{2u}$ at 46~cm$^{-1}$, (L2) $A_{2g}$ ($\Gamma_2^+$, rigid-unit mode) at 53~cm$^{-1}$, (L3) $E_{2g}$ at 59~cm$^{-1}$ and (L4) $A_{1g}$ at 71~cm$^{-1}$. For each mode, three views are shown: oblique, along $c$, and side. The longest arrow corresponds to a displacement of 2~\AA. These modes lie at the upper edge of the spin-wave band (6.2~meV); acoustic and low optical branches elsewhere in the Brillouin zone also contribute to the powder-averaged intensity at low energy.}
\end{figure}

\end{document}